\documentclass[a4paper,11pt]{article}
\pdfoutput=1
\usepackage{jcappub} 
\usepackage[T1]{fontenc} 

\usepackage{graphicx}
\usepackage{multirow}
\usepackage{subfigure}
\usepackage{booktabs}
\newcommand{\be}{\begin{equation}}
\newcommand{\ee}{\end{equation}}
\newcommand{\bea}{\begin{eqnarray}}
\newcommand{\eea}{\end{eqnarray}}

\DeclareRobustCommand{\Fig}[1]{Fig.~\ref{#1}}

\DeclareRobustCommand{\r}[1]{{\rm #1}}

\title{\boldmath 21-cm constraints on spinning primordial black holes}

\author[a,b]{Junsong Cang,}
\author[a]{Yu Gao,}
\author[c,d]{Yin-Zhe Ma}

\affiliation[a]{Key Laboratory of Particle Astrophysics, Institute of High Energy Physics, Chinese Academy of Sciences, Beijing, 100049, China}
\affiliation[b]{School of Physical Sciences, University of Chinese Academy of Sciences, Beijing, 100049, China}
\affiliation[c]{School of Chemistry and Physics, University of KwaZulu-Natal, Westville Campus, Private Bag X54001, Durban, 4000, South Africa}
\affiliation[d]{NAOC–UKZN Computational Astrophysics Centre (NUCAC), University of KwaZulu-Natal, Durban, 4000, South Africa}

\emailAdd{cangjs@ihep.ac.cn}
\emailAdd{gaoyu@ihep.ac.cn}
\emailAdd{ma@ukzn.ac.za}

\abstract{
Hawking radiation from primordial black holes (PBH) can ionize and heat up neutral gas during the cosmic dark ages, leaving imprints on the global 21-cm signal of neutral hydrogen. We use the global 21-cm signal to constrain the abundance of spinning PBHs in mass range of $[2 \times 10^{13}, 10^{18}]$ grams. We consider several extended PBH distribution models. Our results show that 21-cm can set the most stringent PBH bounds in our mass window. Compared with constraints set by {\it Planck} cosmic microwave background (CMB) data, 21-cm limits are more stringent by about two orders of magnitudes. PBHs with higher spin are typically more strongly constrained. Our 21-cm constraints for the monochromatic mass distribution rule out spinless PBHs with initial mass below $1.5 \times 10^{17}\ \r{g}$, whereas extreme Kerr PBHs with reduced initial spin of $a_0=0.999$ are excluded as the dominant dark matter component for masses below $6 \times 10^{17}\ \r{g}$. We also derived limits for the log-normal, power-law and critical collapse PBH mass distributions.
}
\begin{document}
\maketitle

\section{Introduction}

The primordial black hole (PBH) is a type of hypothetical black holes (BH) that can be produced in the very early Universe due to the gravitational collapse of highly over-dense regions~\cite{Hawking:1971ei,Carr:1975qj,1978MNRAS.184..721C,Khlopov:2008qy}.
In addition to the connection with the theories of the early Universe,
PBHs are considered as a dark matter (DM) component in numerous literature~\cite{Carr:2016drx,Kashlinsky:2019kac,Carr:2020xqk,Carr:2020gox,Dasgupta:2019cae,Laha:2020vhg,Ray:2021mxu}.
Compared with astrophysical BHs, PBHs can have masses ranging from $10^{-5}$ grams to beyond $10^{12}$ solar masses~\cite{Carr:2020erq,Carr:2020xqk}.
These BHs can be detected through various techniques~\cite{Carr:2020xqk,Laha:2019ssq,Dasgupta:2019cae,Laha:2020ivk,Wang:2020uvi,Mittal:2021egv,Mena:2019nhm,Villanueva-Domingo:2021cgh,Hektor:2018qqw,Halder:2021rbq,Halder:2021jiv,Halder:2021rrh,Carr:2020gox,Carr:2020erq,Laha:2020vhg,Ray:2021mxu,Yang:2020zcu}, 
ranging from gravitational wave~\cite{Chen:2019xse} and lensing~\cite{Griest:2013esa} to accretion~\cite{Ali-Haimoud:2016mbv,Tashiro:2012qe,Yang:2021idt,Yang:2021agk}.

PBHs lighter than $10^{18}$ grams emit Hawking radiation above keV, 
which can efficiently ionize and heat up the inter-galactic medium (IGM).
The ionizing effect of PBH radiation increases number density of free electrons,
which subsequently scatter off cosmic microwave background (CMB) photons~\cite{Padmanabhan:2005es}, and PBHs can be constrained by the measurements of anisotropy spectra of CMB~\cite{Cang:2020aoo,Poulter:2019ooo,Stocker:2018avm,Acharya:2020jbv,Clark:2016nst,Poulin:2016anj}. Also, IGM gas can be heated by PBH radiation, leading to a correction in the expected 21-cm signal from neutral hydrogen~\cite{Mack:2008nv,Clark:2018ghm,Yang:2020egn},
which arises from hyperfine energy split between parallel and anti-parallel spin states of ground state hydrogen due to interaction of the magnetic moments between proton and electron~\cite{Pritchard:2011xb}.
Ignoring spatial inhomogeneity, the mean 21-cm brightness temperature in $\Lambda{\r{CDM}}$ cosmology is given by~\cite{Pritchard:2011xb,Mesinger:2010ne},
\be
T_{21}
\approx
27
x_{\r{HI}}
\left(
\frac{\Omega_{\r{b}}h^2}{0.023}
\right)
\left(
\frac{0.15}{\Omega_{\r{m}}h^2}
\frac{1+z}{10}
\right)^{1/2}
\left(
\frac{T_{\r{S}}-T_{\r{CMB}}}{T_{\r{S}}}
\right)
\r{mK}
\label{fdgfhejwuy7}
\ee
where $x_{\r{HI}}$ and $T_{\r{CMB}}$ are neutral hydrogen fraction and CMB temperature.
$T_{\r{S}}$ is the spin temperature that characterizes the occupation numbers on the hydrogen atom's split hyperfine states.
After gas and the CMB decouple after around $z\sim 200$, $T_{\rm S}$ evolves separately from $T_{\rm CMB}$. In particular in the early reionization era, $T_S$ becomes tightly coupled to gas temperature $T_{\r{K}}$ at around redshift $z=20$ due to Wouthuysen-Field effect~\cite{Pritchard:2011xb,Mesinger:2010ne,Kuhlen:2005cm,Ciardi:2003hg,1958PIRE...46..240F}. The PBH radiation's potential heating effect on IGM gas temperature $T_{\rm K}$ would also raise $T_{\rm S}$, causing a reduction in the expected 21-cm signal strength in comparison with that in vanilla $\Lambda$CDM.

The first detection of the 21-cm signal was claimed by the EDGES collaboration~\cite{Bowman:2018yin},
which measured a $T_{21}$ trough centered around redshift $z=17$:
$
T_{21}=-500^{+200}_{-500} \  {\r{mK}}
$.
Although the timing of this absorption profile is thought to be consistent with astrophysical expectations, 
the amplitude of the signal is deeper than the largest predictions by more than a factor of two~\cite{Cohen:2016jbh}.
Such deep absorption trough might be possible if excess radio background other than CMB were present~\cite{Feng:2018rje,Ewall-Wice:2018bzf,Ewall-Wice:2019may,Jana:2018gqk,Fialkov:2019vnb},
or if the gas temperature at this redshift were colder than 3.2 K~\cite{Bowman:2018yin},
which can be achieved via the cooling effect of baryon-DM scattering~\cite{Tashiro:2014tsa,Munoz:2015bca,Barkana:2018lgd,Bhatt:2019lwt}.
In contrast, gas heating due to Hawking radiation raises $T_{\rm K}$ and the spin temperature $T_{\r{S}}$ hence decreases the $T_{21}$ amplitude,
therefore Hawking radiation of PBH cannot explain the depth of EDGES signal.
Nonetheless one can obtain very stringent limits on PBH abundance by requiring that the PBH heating does not wipe out the 21-cm signal~\cite{Clark:2018ghm,Yang:2020egn,Natwariya:2021xki}.

Using the gas temperature upper limit inferred from a $T_{21}$ upper bound consistent with EDGES result,
in this work we constrain PBHs in the $[2 \times 10^{13}, 10^{18}]$ g mass window with several mass distributions.
PBHs are normally assumed to have low spins~\cite{Chiba:2017rvs,Mirbabayi:2019uph,DeLuca:2019buf},
yet in some schemes it's also possible for them to have high 
spins~\cite{Dasgupta:2019cae,Khlopov:1980mg,Harada:2017fjm,Cotner:2018vug,Kokubu:2018fxy,He:2019cdb,Bai:2019zcd,Cotner:2019ykd,Arbey:2019jmj,Flores:2021tmc}.
Therefore in addition to the conventional Schwarzschild PBHs,
we will also explore the parameter space of spinning Kerr PBHs.
Although PBHs with initial mass below $10^{15}$ g would have evaporated away,
they can still leave imprints on cosmic ionization and thermal history~\cite{Stocker:2018avm,Acharya:2020jbv,Mack:2008nv}.
We also addressed several technical complications involving PBHs lighter than $10^{15}$ g.
The first is that the emission of unstable standard particles, 
which decay or hadronize almost immediately after being emitted,
will produce a cascade of secondary particles that carry on energy in the form of long-lived $\gamma$, $e^{\pm}$ and $\nu$~\cite{MacGibbon:1990zk}.
Another complication is that the mass variation of PBHs lighter than $10^{15}$ g can no longer be ignored,
which means that the PBH mass distribution would also evolve during the dark ages.

We discuss the particle spectra of Hawking evaporation and the evolution of mass and spin distribution in Sec.~\ref{dsghfjdsfegf}.
Sec.~\ref{dsfguhsivudy} examines the impacts of PBHs on the evolution of cosmic gas temperature and ionization level.
Our results are presented in Sec.~\ref{sdgsfhgvhdhh45rtgh} and we conclude in Sec.~\ref{sdgsfhgvhdhh45rtgh_2}.
We assume the spatially-flat $\Lambda \r{CDM}$ cosmology with the relevant parameters set by {\it{Planck}} 2018 results~\cite{Planck:2018vyg}:
$h=0.6766$,
$\Omega_{\r{b}} h^2=0.02242$,
$\Omega_{\r{c}} h^2=0.11933$,
$\tau=0.0561$,
${\r{log}}(10^{10}A_{\r{s}})=3.047$,
$n_{\r{s}}=0.9665$.

\section{Evaporating Black Holes}
\label{dsghfjdsfegf}
\subsection{Particle Spectra}

The temperature of Hawking radiation emitted by a Kerr black hole is~\cite{PhysRevD.13.198,PhysRevD.14.3260,Arbey:2019mbc,Dasgupta:2019cae,Arbey:2021mbl}
\be
T_{\r{PBH}}
=
\frac{1}{4 \pi M}
\frac{\sqrt{1-a^2}}{1+\sqrt{1-a^2}}
=
2.12
\times
\frac{10^3\r{g}}{M}
\frac{\sqrt{1-a^2}}{1+\sqrt{1-a^2}}
\r{GeV}
,
\ee
where $M$ is PBH mass,
$a$ is the reduced dimensionless spin parameter, defined as
\be
a
\equiv
J/M^2
\in
[0,1]
,
\ee
and $J$ is the angular momentum of the BH.
The primary particle emission spectra of a BH is given by~\cite{Hawking:1975vcx,PhysRevD.14.3260,MacGibbon:1990zk,Arbey:2019mbc,Arbey:2021mbl}
\be
\left[
\frac{\r{d}N^{\alpha}}{\r{d}E\r{d}t}
\right]_{\r{pri}}
=
\frac{1}{2 \pi}
\sum_{\r{dof}}
\frac{\Gamma^{\alpha}}
{\r{e}^{E'/T_{\r{PBH}}} - (-1)^{2s}}
,\ 
E'
=
E-
\frac{m}{2M}
\frac{a}{1+\sqrt{1-a^2}}
,
\ee
here the subscript `pri' denotes primary emission,
$\alpha$ labels particle species,
$s$ is the spin of the particle,
$\Gamma^{\alpha}$ is the greybody factor~\cite{PhysRevD.14.3260,Arbey:2019mbc,MacGibbon:1990zk,Arbey:2021mbl}.
The sum is over the total multiplicity of the particle as well as the angular momentum $l$ 
and its $z$-component $m \in \{-l,\ ...,\ l\}$.

The prompt particle spectra
$\left({\r{d}N^{\alpha}}/{\r{d}E\r{d}t}\right)$
for stable particle species,
which contains both primary and secondary emissions,
is computed as (see also Ref.~\cite{Arbey:2019mbc,Arbey:2021mbl}),
\be
\left[
\frac{\r{d}N^{\alpha}}{\r{d}E\r{d}t}
\right]
(M,a,E)
=
\left[
\frac{\r{d}N^{\alpha}}{\r{d}E\r{d}t}
\right]_{\r{pri}}
(M,a,E)
+
\frac{1}{2}
\sum_{\alpha'}
\int
dE'
\frac{dN^{\alpha}}{dE}(E,E')
\left[
\frac{\r{d}N^{\alpha'}}{\r{d}E\r{d}t}
\right]_{\r{pri}}
(M,a,E')
,
\label{9dsj32ghjhg}
\ee
where ${\r{d}N^{\alpha}}/{\r{d}E}$ is the energy spectra of $\alpha$ produced by decay/hadronization of a pair of unstable $\alpha '$ particles with center of mass energy of $2E'$,
$\alpha '$ runs through all directly emitted standard model particles except for $(e^{\pm},\gamma, \nu)$.

We use the {\tt{BlackHawk}} package~\cite{Arbey:2019mbc,Arbey:2021mbl} to calculate the particle spectra 
$\left({\r{d}N^{\alpha}}/{\r{d}E\r{d}t}\right)$
and evolution of PBH mass $M$ and spin $a$.
Due to Hawking radiation,
a PBH will lose its mass at a rate given by~\cite{Arbey:2019mbc,PhysRevD.14.3260,Dong:2015yjs,Arbey:2021mbl},
\be
\frac{\r{d}M}{\r{d}t}
=
-
\frac{f(M)}
{M^2}.
\label{dsjhghfsvhc2e}
\ee
Here $f$ is the Page factor~\cite{PhysRevD.14.3260},
\be
f
\equiv
-M^2
\frac{\r{d}M}{\r{d}t}
=
M^2
\sum_{\alpha}
\int_0^{\infty}
\r{d} E
\ 
E
\left[
\frac{\r{d}N^{\alpha}}{\r{d}E\r{d}t}
\right]_{\r{pri}}
,
\ee
and the sum is over all standard model particle species.
For the Schwarzschild PBHs $(a=0)$ in particular,
the Page factor can be written as~\cite{MacGibbon:1991tj,Stocker:2018avm}
\bea
f
&=&
5.34 \times 10^{25}
\mathcal{F}
\ 
\r{g}^3
/\r{s}
,
\label{dhsfjsi65}
\\
\nonumber
\\
\mathcal{F}
&=&
\sum_{\alpha}
g^{\alpha}
\omega^{\alpha}
\r{exp}
\left(
-M/M^{\alpha}
\right)
\ 
\in
\ 
[1,16.4]
.
\label{fdsfjhefghvs}
\eea
Here $g^{\alpha}$ is the internal degrees of freedom of particle $\alpha$,
$\omega^{\alpha}$ and $M^{\alpha}$ are relativistic emission fraction and characteristic BH mass for each particle species, 
whose values have been tabulated in Refs.~\cite{Stocker:2018avm}.
In general, smaller PBHs evaporates more violently than heavier ones,
a Schwarzschild PBH with $M<7.9 \times 10^{14}\r{g}$ would have completely evaporated before the present day.
In $[2 \times 10^{13}, 10^{18}]$ g mass range considered here, such mass loss can cause non-negligible modification to BH mass distributions and energy injection history.

\subsection{Evolving Mass and Spin Distributions}
When mass loss and spin-down is significant during the age of the Universe, the PBH mass and spin evolution with redshift need to be taken care of properly. 
Ignoring mergers, the number of PBHs are conserved prior to their final complete evaporation. 
The mass distribution of PBHs can be described through their mass density $\rho$ and number density $n$ at comoving frame,
\be
\Psi(M,t) 
\equiv
\frac{1}
{\rho_0}
\frac{
{\rm{d}}{\rho}
}
{{\rm{d}}M}
,
\label{shffw67tyh}
\ee
\be
\Phi(M,t) 
\equiv
\frac{1}
{n_0}
\frac{
{\rm{d}}{n}
}
{{\rm{d}}M}
,
\label{MF_DEF2}
\ee
where $\rho_0$ and $n_0$ are initial {\it comoving} BH mass and number density at some early time $t=t_0$,
\be
\rho_0
=
f_{\rm{PBH}}
\rho_{\r{c}}
,
\label{dsfefhjedwjk_2}
\ee
here $\rho_{\r{c}}$ is the comoving cold dark matter density,
$f_{\r{PBH}} \equiv \rho_0/\rho_{\r{c}}$ is the initial fraction of dark matter made of PBH.

We assume that $t_0 \ll \tau_{\r{min}}$,
where $\tau_{\r{min}} \sim 5 \times 10^4\ \r{yrs}$ is the lifetime of the most short-lived PBH considered here.
Both $\Psi(M,t)$ and $\Phi(M,t)$ are normalized to unity at $t_0$, 
however they do not stay normalized if a fraction of PBH evaporate away at $t>t_0$.
One can obtain $\Phi(M,t)$ from $\Psi(M,t)$ (and vice versa) by
\be
\Delta n
=
n_0
\int^{M''}_{M'}
\r{d}M
\Phi(M,t)
=
\rho_0
\int^{M''}_{M'}
\r{d}M
\frac{\Psi(M,t)}{M}
,
\ee
which gives,
\be
\Phi
=
C
\frac{\Psi}{M}
,
\ \ 
C^{-1}
=
\int_0^{\infty}
\r{d}M
\frac{\Psi(M,t_0)}
{M},
\ee
with the normalization factor $C$ set by the normalization conditions for $\Psi$ and $\Phi$ at $t_0$,
\be
1=
\int_0^{\infty}
\r{d}M \Phi(M,t_0)
=
C
\int_0^{\infty}
\r{d}M
\frac{\Psi(M,t_0)}
{M}
.
\ee

Since $\Psi$ and $\Phi$ are essentially equivalent, hereafter we will use $\Psi$ to describe PBH mass distribution.
In the simplified monochromatic approximation,
in which all PBHs have the same mass,
\be
\Psi(M,t_0)
=
\delta_{\r{D}}(M-M_0)
,
\label{fjdhfgevf675rtfg}
\ee
where $M_0$ indicates the initial mass of BH at $t_0$.
This monochromatic model is a convenient approximation,
and many PBH formation theories tend to favor extended distributions~\cite{Dolgov:1992pu,Carr:2017jsz,Green:2016xgy,Yokoyama:1998xd,Niemeyer:1999ak,Bellomo:2017zsr,Pi:2017gih},
therefore we also consider the following three theoretically motivated extended mass distribution scenarios:

\begin{itemize} 
\item Log-normal model~\cite{Dolgov:1992pu,Carr:2017jsz,Green:2016xgy}
\end{itemize}
\bea
\Psi(M,t_0)
=
\frac{1}
{\sqrt{2 \pi} \sigma M}
{\rm{exp}}
\left(
-
\frac{({\rm{log}}[M/M_\r{c}])^2}
{2 \sigma^2}
\right)
,
\label{fdsdsahfg54esdf1qw}
\eea

\begin{itemize}
\item Critical Collapse model~\cite{Yokoyama:1998xd,Niemeyer:1999ak,Carr:2017jsz}
\end{itemize}
\be
\Psi(M,t_0)
=
\frac{3.2}
{M_{\r{c}}}
\left(\frac{M}{M_{\r{c}}}\right)^{2.85}
{\rm{exp}}
\left[{-\left(\frac{M}{M_{\r{c}}}\right)^{2.85}}\right]
,
\label{dsfejh76tg}
\ee

\begin{itemize}
\item Power-law model~\cite{Carr:2017jsz} 
\end{itemize}
\be
\Psi(M,t_0)
=
\frac{\gamma}
{M_{\r{max}}^{\gamma}-M_{\r{min}}^{\gamma}}
M^{\gamma -1}
,
\ 
M \in [M_{\rm{min}}, M_{\rm{max}}]
.
\label{MdsgfF_3}
\ee

Here the power-law index $\gamma$ is related to the equation of state parameter $\omega$ during PBH formation by
$\gamma = -2 \omega/(1+\omega)$,
therefore for post-inflation epochs $0<|\gamma| \le 1$.
PBH production at matter dominated era with $\gamma = 0$ requires special treatment (e.g. see~\cite{Carr:2017edp})
and is not considered here.

Note that Eqs. (\ref{fdsdsahfg54esdf1qw}) - (\ref{MdsgfF_3}) are all given at $t=t_0$,
their evolved forms can be tracked by imposing that the BH numbers are conserved (until before the smallest BH evaporate away),
\be
\Delta n
=
\rho_0
\int^{M_0''}_{M_0'}
{\r{d}}M_0
\ 
\frac{\Psi(M_0,t_0)}{M_0}
=
\rho_0
\int^{M''}_{M'}
{\r{d}}M
\ 
\frac{\Psi(M,t)}{M}
,\ 
M'>0
\label{DDGJHJ_2}
\ee
which gives
\be
\Psi(M,t)
=
\Psi[M_0,t_0]
\times
\frac{{\r{d}}M_0}{{\r{d}}M}
\frac{M}{M_0}
,
\label{dksgdshj_1}
\ee
where $M_0(M,t)$ indicates the initial mass of a BH with mass $M$ at $t$.
Analogously one can start from
\be
\Delta n
=
n_0
\int^{M_0''}_{M_0'}
{\r{d}}M_0
\cdot
\Phi(M_0,t_0)
=
n_0
\int^{M''}_{M'}
{\r{d}}M
\cdot
\Phi(M,t)
,\ 
M'>0
,
\label{DDGJHJ}
\ee
to show that
\be
\Phi(M,t)
=
\Phi[M_0,t_0]
\times
\frac{{\r{d}}M_0}{{\r{d}}M}
.
\label{dksgdshj_2}
\ee

In addition to Hawking radiation, there could also be other physical processes that can change the mass and spin of PBHs,
most notably BH mergers~\cite{Fishbach:2017dwv,Doctor:2021qfn,Fakhry:2020plg,Fakhry:2021tzk} and accretion~\cite{LKrol:2021pau}.
It's possible that PBH spin can also have extended distributions~\cite{Arbey:2019mbc,Flores:2021tmc,LKrol:2021pau,Fishbach:2017dwv,Doctor:2021qfn,Nelson:2019czq,Jaraba:2021ces,Harada:2016mhb,Arbey:2021mbl}, 
but these complicated scenarios are beyond the scope of this paper, 
therefore we will adopt the same strategy as Refs.~\cite{Dasgupta:2019cae,Laha:2020vhg,Ray:2021mxu} and restrict our discussions to PBHs born with the same initial spin.

Equation (\ref{DDGJHJ_2}) is equivalent to the requirement that Hawking radiation does not change mass ordering of PBHs,
such that any BH initially heavier than $M_0'$ will always be present provided that $M_0'$ does not vanish at $t$.
This is true for Schwarzschild PBHs, for which heavier PBHs always evaporate at a slower rate.
For Kerr PBHs however, 
it's more complicated as a BH may lose energy faster than a less massive one of lower spin.
In practice, one may always bin the PBH spin distribution and sum up the evolved contributions from each bin later. 
Thus we can consider BHs of some fixed initial spin $a_0$.
For each fixed initial spin, we numerically simulated the mass evolution $M_0$ values in the $[2 \times 10^{13}, 10^{18}]$ g range,
and we find that the ordering of evolved mass is maintained in the redshift range of interest ($z \in [11,2.7 \times 10^3]$).
Therefore Eq. (\ref{dksgdshj_1}) also holds for PBHs with a common initial spin, and we will use it to study the evolved distribution for both Schwarzschild PBHs and Kerr BHs with a certain initial spin.

\begin{figure*}[h]
\centering
\subfigbottomskip=-200pt
\subfigcapskip=-7pt
\subfigure{\includegraphics[width=8cm]{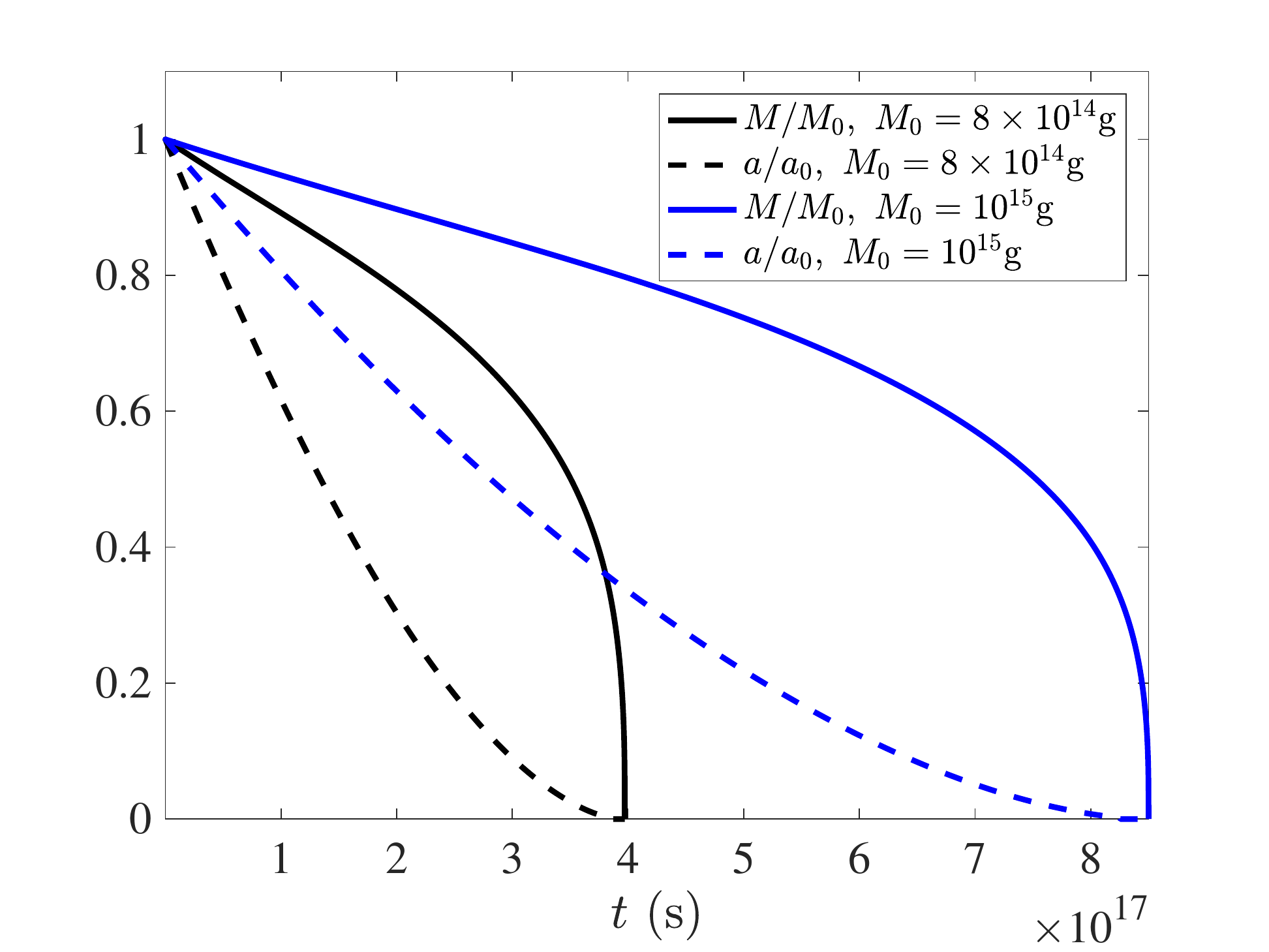}\includegraphics[width=8cm]{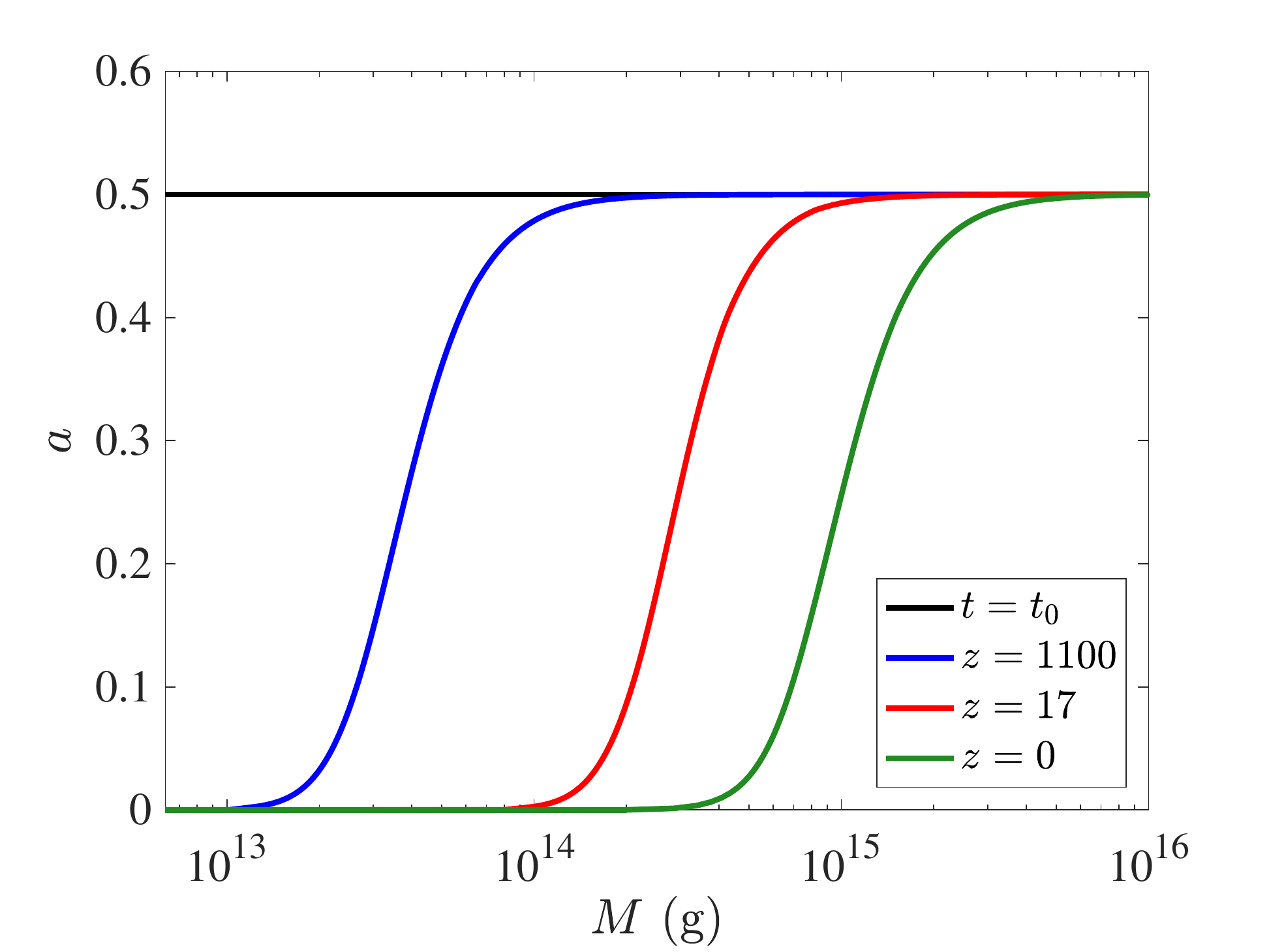}}
\caption{
{\it{Left}}:
Mass (solid) and spin (dashed) evolution of two PBHs with initial masses of $8 \times 10^{14}\ \r{g}$ (black) and $10^{15}\ \r{g}$ (blue).
{\it{Right}}:
Evolved spin for survived PBH masses at redshifts of 1100  (blue), 17 (red) and 0 (green).
All PBHs are assumed to have same initial spin of $a_0=0.5$ in both panels.
}
\label{adsfsdh654er}
\end{figure*}

Hawking radiation also cause a BH to gradually lose its spin at a rate given by (see also~\cite{Arbey:2019mbc,PhysRevD.14.3260,Dong:2015yjs,Arbey:2021mbl}),
\bea
\frac{\r{d}a}{\r{d}t}
&=&
a
\frac{2f-g}{M^3}
,
\\
g
&\equiv&
-
\frac{M}{a}
\frac{\r{d}J}{\r{d}t}
=
\frac{M}{a}
\sum_{\alpha}
\int_0^{\infty}
\r{d} E
\ 
m
\left[
\frac{\r{d}N^{\alpha}}{\r{d}E\r{d}t}
\right]_{\r{pri}}
.
\eea
An example of mass and spin evolution as computed by {\tt BlackHAWK} is shown in the left panel of Fig.~\ref{adsfsdh654er}.
Typically smaller PBHs also lose their spin faster than heavier ones,
therefore for PBHs with the same initial spin,
as Hawking radiation progresses,
PBH mass and spin evolution will become correlated.
\begin{figure*}[htp]
\centering
\subfigbottomskip=-200pt
\subfigcapskip=-7pt
\subfigure{\includegraphics[width=8cm]{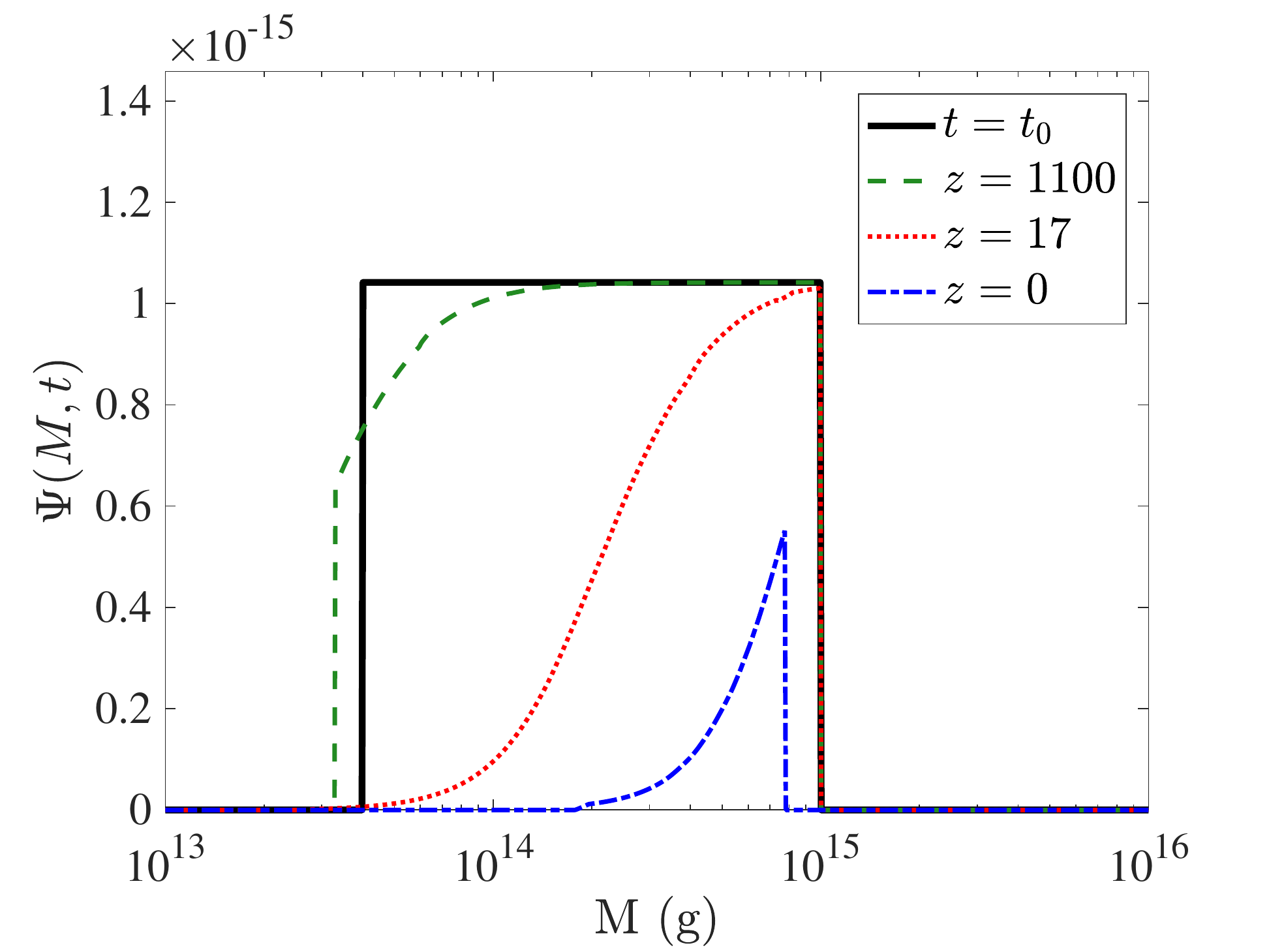}\includegraphics[width=8cm]{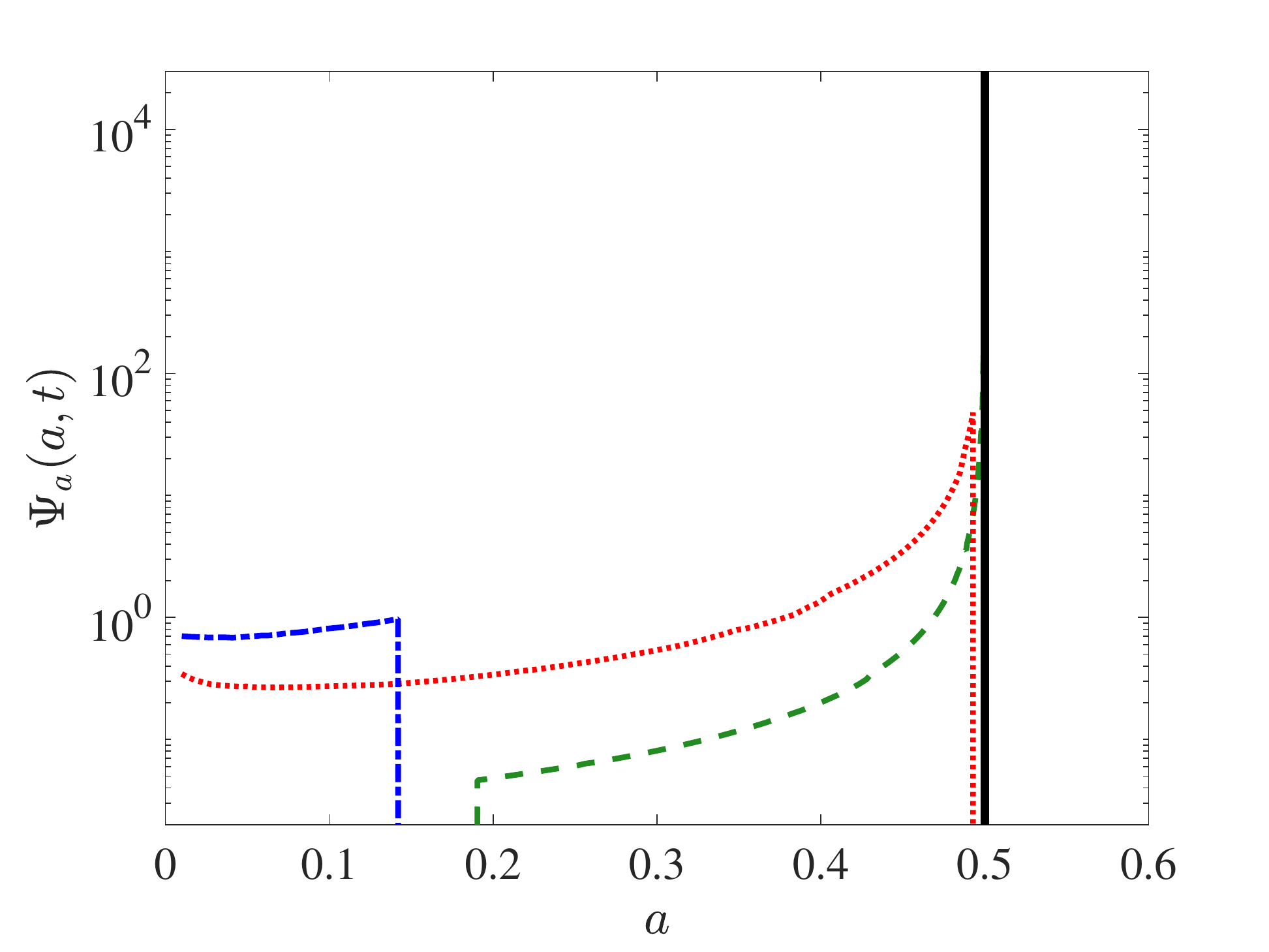}}
\caption
{
Evolution of a uniform box-like power-law mass distribution ({\it left}) with $\gamma=1,\ M_{\r{min}}=4 \times 10^{13} \r{g},\ M_{\r{max}}=10^{15} \r{g}$,
the corresponding spin distributions are shown in the {\it right} panel.
The black solid, green dashed, red dotted, blue dot-dashed lines correspond to redshifts of $\infty\ (t=t_0)$, $1100$ (recombination), $17$ (EDGES redshift) and $0$ (present) respectively.
All PBHs are assumed to have the same initial spin of $a_0=0.5$,
legend applies to both panels.
}
\label{MassFunction}
\end{figure*}
The right panel in Fig.~\ref{adsfsdh654er} shows the evolution of spin-mass relation for an initial spin of $a_0=0.5$.
For extended mass distributions,
this correlation will cause PBH spins to spread from a monochromatic initial distribution to an extended distribution,
which itself is related to the mass distribution simply by,
\be
\Psi_a(a,t)
\equiv
\frac{1}
{\rho_0}
\frac{
{\rm{d}}{\rho}
}
{{\rm{d}}a}
=
\Psi[M(a),t]
\frac{\r{d}M}
{\r{d}a}
,
\label{dshifihe398ui}
\ee
where $\Psi_a(a,t)$ is the spin distribution.
The monochromatic initial spin distribution as considered here is given by
\be
\Psi_a(a,t_0) 
=
\delta_{\r{D}}(a-a_0),
\ee

In Fig.~\ref{MassFunction} we show the evolution of a uniform box-like mass distribution and the corresponding spin distribution.
The black solid lines show the initial distributions,
and their evolved forms given by Eqs.~(\ref{dksgdshj_1}) and (\ref{dshifihe398ui}) are shown in green dashed, red dotted and blue dot-dashed lines.
Since the fractional mass and spin loss of even the smallest PBH considered in the figure only becomes noticeable after $z=1100$,
the evolution in $\Psi(M,t)$ and $\Psi_a(a,t)$ remains negligible until the recombination epoch,
and the numerical integration of the green lines in both panels consistently reveals that 
$\rho/\rho_0=  \int \r{d} a \Psi_a(a,t)=\int \r{d}M \Psi(M,t) > 0.99$,
which suggests that at $z=1100$,
more than 99\% of the initial PBH mass density $\rho_0$ remain untouched by evaporation.

\section{PBH imprints on IGM}
\label{dsfguhsivudy}

Radiation from evaporating PBHs propagate through the Universe and gradually damp their energy into the IGM through occasional collisions with the intergalactic gas. Consequent impact on the CMB and 21-cm mainly occurs via enhanced ionization fraction and gas temperature. In this section, we will calculate the energy injection rate due to PBH radiation for a few characteristic PBH mass distributions, and their energy deposit rate into the IGM. Such energy deposit sources will modify the evolution of IGM temperature $T_{\r{K}}$ and ionization level $x_{\r{e}} \equiv n_{\r{e}}/n_{\r{H}}$ during the cosmic dark age.

In the scope of this paper, 
energy released by evaporating PBHs is absorbed by the IGM mainly through three deposition channels:
IGM heating (Heat), hydrogen ionization (HIon) and excitation (Ly$\alpha$).
In presence of these extra energy deposition processes, 
the modification to the standard evolution equations for $T_{\r{K}}$ and $x_{\r{e}}$ take the form~\cite{Slatyer:2016qyl,Liu:2016cnk,Cang:2020exa,Cang:2020aoo},

\bea
\frac{{\rm{d}}T_{\rm{K}}}
{{\rm{d}} t}
& = &
\left[
\frac{{\rm{d}}T_{\rm{K}}}
{{\rm{d}} t}
\right]_0 
+ 
\frac{2}{3n_{\r{H}}(1+f_{\rm{He}}+x_{\rm{e}})}
\left[
\frac{{\rm{d}}E}
{{{\rm{d}}V}{\rm{d}}t}
\right]_{\rm{dep,Heat}} ,
\label{dsfeffbshfdhvb761}
\\
\nonumber
\\
\frac{{\rm{d}}x_{\rm{e}}}
{{\rm{d}} t}
& = &
\left[
\frac{{\rm{d}}x_{\rm{e}}}
{{\rm{d}} t}
\right]_0 
+
\frac{1}{n_{\rm{H}}(z)E_{\r{i}}}
\left[\frac{{\rm{d}}E}{{\rm{d}}V{\rm{d}}t}\right]_{{\rm{dep,HIon}}}
+
\frac{1-C}{n_{\rm{H}}(z)E_\alpha}
\left[\frac{{\rm{d}}E}{{\rm{d}}V{\rm{d}}t}\right]_{{\rm{dep,Ly\alpha}}} .
\label{dsfeffbshfdhvb76}
\eea
Here $E_{\r{i}}=13.6 \ \r{eV}$ and $E_{\alpha}=10.2\  \r{eV}$. The first term with subscript 0 denotes for the standard evolution equation (see Refs.~\cite{Ali-Haimoud:2010hou,Seager:1999bc}) without PBH radiation injection. 
$
\left[
{{\rm{d}}E}/
{{{\rm{d}}V}{\rm{d}}t}
\right]_{\rm{dep,c}}
$
is the energy deposition rate per unit volume (hereafter dubbed deposition rate for simplicity) through absorption channel $\r{c} \in [\r{HIon}, \r{Ly\alpha}, \r{Heat}]$.
$f_{\rm{He}}$ is the helium fraction by number of nuclei,
$C$ is Pebble's C factor~\cite{Seager:1999bc,Slatyer:2016qyl,Liu:2016cnk} which describes the probability for an excited $n=2$ hydrogen atom to transit back to ground state before being ionized. Eq.~\ref{dsfeffbshfdhvb761} and ~\ref{dsfeffbshfdhvb761} are implemented in our customized {\tt HyRec} package to compute the $T_{\rm K}$ and $x_e$ evolution with PBH radiation injection contributions.

Since $e^{\pm}$ and $\gamma$ make up the majority of emitted stable particles with sufficient electromagnetic interaction with the IGM,
one can calculate the energy deposition rate by tracking the IGM interaction of radiated $e^{\pm}$ and $\gamma$,
whose energy deposition processes at redshift $z$ can be described by a transfer function $\mathcal{T}^{\alpha}_{\r{c}}(z,E,z')$~\cite{Slatyer:2015kla} from the radiation energy injected at an earlier redshift $z'$.
For a particle $\alpha \in [\gamma,e^{\pm}]$ injected at $z'$ with energy $E$, 
$\mathcal{T}^{\alpha}_{\r{c}}(z,E,z')$ gives the fraction of $E$ absorbed into channel $\r{c}$ during unit $-\r{log}(1+z)$ interval.
For a generic particle injection history,
the relevant deposition rate is given by (see also~\cite{Slatyer:2012yq,Slatyer:2015kla}),
\be
\begin{aligned}
\left[
\frac{{\rm{d}}E}
{{\rm{d}}V{\rm{d}}t}
\right]_{\rm{dep,c}}
(z)
& =
\frac{\r{d}x}{\r{d}t}
\int
\r{d} t'
\sum_{\alpha = \gamma, e^{\pm}}
\left[
\int \r{d}E
\ 
E
\mathcal{T}^{\alpha}_{\r{c}}(z,E,z')
\mathcal{I}^{\alpha}(E,z')
\frac{\r{d}V'}{\r{d}V}
\right]
\\
&=
(1+z)^3 H(z)
\int
\frac
{\r{d} z'}
{(1+z')^4 H(z')}
\sum_{\alpha = \gamma, e^{\pm}}
\left[
\int \r{d}E
\ 
E
\mathcal{T}^{\alpha}_{\r{c}}(z,E,z')
\mathcal{I}^{\alpha}(E,z')
\right]
,
\end{aligned}
\label{Dep_EFF_EQ}
\ee
where $x \equiv -\r{log}(1+z)$, and in the second line we used relations $\r{d}t' = \r{d}x'/H(z')$, 
$\r{d}V \propto (1+z)^{-3} $ and $\r{d}V' \propto (1+z')^{-3} $.
$\mathcal{I}^{\alpha}$ is the differential particle injection rate per unit volume,
\be
\mathcal{I}^{\alpha} \equiv 
\frac
{\r{d} N^{\alpha}}
{\r{d}E \r{d}V \r{d}t},
\ee
we will elaborate on this term in next two subsections.

\subsection{Monochromatic mass distribution}

Assuming a monochromatic PBH mass distribution $\delta(M-M_0)$, and a given initial spin $a_0$,
the particle injection history $\mathcal{I}^{\alpha}$ in Eq. (\ref{Dep_EFF_EQ}) takes the form,
\be
\mathcal{I}^{\alpha,\delta}
(M_0,a_0,E,z)
=
\left[
\frac{\r{d}N^{\alpha}}{\r{d}E\r{d}t}
\right]
(M,a,E)
\cdot
n_{\r{PBH}}
\times
\Theta
\left[
\tau_{\r{PBH}}(M_0,a_0)-t
\right]
\label{dsfghw232}
\ee
where 
$
{\r{d}N^{\alpha}}/{\r{d}E\r{d}t}
$
is computed through Eq. (\ref{9dsj32ghjhg}),
superscript $\delta$ indicates monochromatic distribution,
$M$ is the mass of $M_0$ at redshift $z$,
$\tau_{\r{PBH}}$ is the lifetime of $M_0$,
$t$ is the age of the Universe at redshift $z$,
$\Theta$ is the Heaviside step function which enforces $\mathcal{I}^{\alpha,\delta}$ to vanish once $M_0$ reaches its' end of life.
$n_{\r{PBH}}$ is PBH number density,
\be
n_{\rm{PBH}} (z)= 
f_{\rm{PBH}}
\frac{\Omega_{\rm{c}} \rho_{\rm{cr}} (1+z)^3}
{M_0}.
\label{n_PBH}
\ee
where $\rho_{\r{cr}}$ is the current critical density of the Universe.
Inserting Eqs. (\ref{dsfghw232},\ref{n_PBH}) into Eq. (\ref{Dep_EFF_EQ}) gives the deposition rate for monochromatic PBHs,
\be
\begin{aligned}
\left[
\frac{{\rm{d}}E}
{{\rm{d}}V{\rm{d}}t}
\right]^\delta_{\rm{dep,c}}
(M_0,a_0,z)
& =
\frac{f_{\r{PBH}} \Omega_{\r{c}} \rho_{\r{cr}} H(z) \left(1+z\right)^3}
{M_0}
\int
\frac
{\r{d} z'}
{(1+z') H(z')}
\\
& \times
\sum_{\alpha = \gamma, e^{\pm}}
\left[
\int \r{d}E
\ 
E
\mathcal{T}^{\alpha}_{\r{c}}(z,E,z')
\left[
\frac{\r{d} N^{\alpha}}
{\r{d} E \r{d} t}
\right]'
\Theta
\left(
\tau_{\r{PBH}}
-
t'
\right)
\right]
,
\end{aligned}
\label{Dep_EFF_EQds}
\ee
here $t'$ is the age of the Universe at redshift $z'$,
$\left[ {\r{d} N^{\alpha}} / {\r{d} E \r{d} t} \right]'$ is the ${\r{d} N^{\alpha}} / {\r{d} E \r{d} t}$ spectra at $z'$,
when the PBH mass and spin would have evolved from $[M_0,\ a_0]$ to $[M,\ a]$. The superscript $\delta$ denotes for the monochromatic distribution.

The energy injection rate per unit physical volume (dubbed injection rate hereafter) from monochromatic PBHs can be written as,
\be
\begin{aligned}
\left[
\frac{{\rm{d}}E}
{{\rm{d}}V{\rm{d}}t}
\right]_{\rm{inj}}^{{\delta}}
(M_0,a_0,z)
& =
\frac{\r{d}E}{\r{d}t}
\cdot n_{\rm{PBH}}
\Theta
\left[
\tau_{\r{PBH}}-t
\right]
.
\end{aligned}
\label{gy76f6f5}
\ee
here ${\r{d}E}/{\r{d}t}$ indicates the power of Hawking radiation,
which is equal to $-{\r{d}M}/{\r{d}t}$ if $a_0=0$.
Eq. (\ref{gy76f6f5}) has a simple analytic form for Schwarzschild PBHs with $M_0 \ge 10^{17}\ \r{g}$,
for which the factor $\mathcal{F}$ in Eq. (\ref{fdsfjhefghvs}) is normalized to unity~\cite{MacGibbon:1991tj,Stocker:2018avm} and therefore
\be
\frac{\r{d}E}{\r{d}t}
=
-\frac{\r{d}M}{\r{d}t}
=
5.34 \times 10^{25}
\left(
\frac{\r{g}}{M}
\right)^2
\r{g}/{\r{s}}.
\label{hiuy8g76f6fr545rt}
\ee
where we have used Eqs. (\ref{dsjhghfsvhc2e},\ref{dhsfjsi65}).
For $M_0 \ge 10^{17}\ \r{g}$, 
the fractional mass loss $\Delta M/M_0$ remains negligible across the entire history of the Universe,
thus one can safely set $M=M_0$ in Eq. (\ref{hiuy8g76f6fr545rt}) and ignore the step function $\Theta \left[ \tau_{\r{PBH}} - t\right]$ in Eq. (\ref{gy76f6f5}),
therefore inserting Eq. (\ref{hiuy8g76f6fr545rt}) into Eq. (\ref{gy76f6f5}) gives,
\be
\left[
\frac{{\rm{d}}E}
{{\rm{d}}V{\rm{d}}t}
\right]_{\rm{inj}}'
=
5.34
\times
10^{25}
({M_0/\r{g}})^{-3}
f_{\r{PBH}}
\Omega_{\r{c}}
\rho_{\r{cr}}
{(1+z)^3}
\ \r{s}^{-1}.
\label{888eygswdgvywy}
\ee
Although $\left[ {{\rm{d}}E}/{{\rm{d}}V{\rm{d}}t} \right]_{\rm{inj}}'$ does not accurately reproduce $\left[ {{\rm{d}}E}/{{\rm{d}}V{\rm{d}}t} \right]^{\delta}_{\rm{inj}}$ if $M_0<10^{17} \ \r{g}$ or $a_0 > 0$,
it is convenient to relate it to the actual deposition rate through an effective deposition efficiency, 
defined as,
\be
f_{\rm{c}}
(z)
\equiv
\left[
\frac{{\rm{d}}E}
{{\rm{d}}V{\rm{d}}t}
\right]_{\rm{dep,c}}^{\delta}
\Big{/}
\left[
\frac{{\rm{d}}E}
{{\rm{d}}V{\rm{d}}t}
\right]_{\rm{inj}}',
\label{kdsfkj887}
\ee
\begin{figure*}[ht] 
\centering
\subfigbottomskip=-200pt
\subfigcapskip=-7pt
\subfigure{\includegraphics[width=5.2cm]{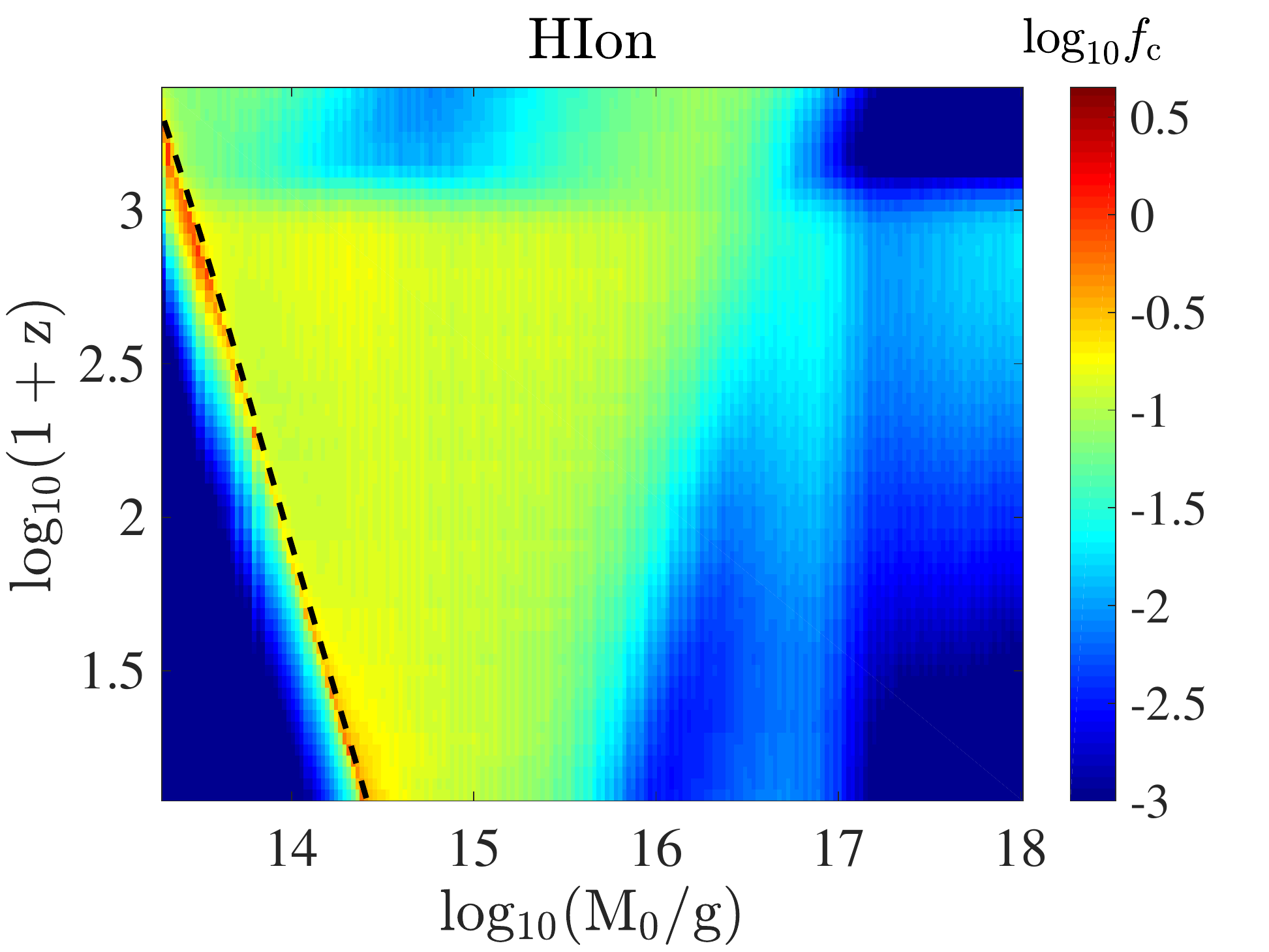} \includegraphics[width=5.2cm]{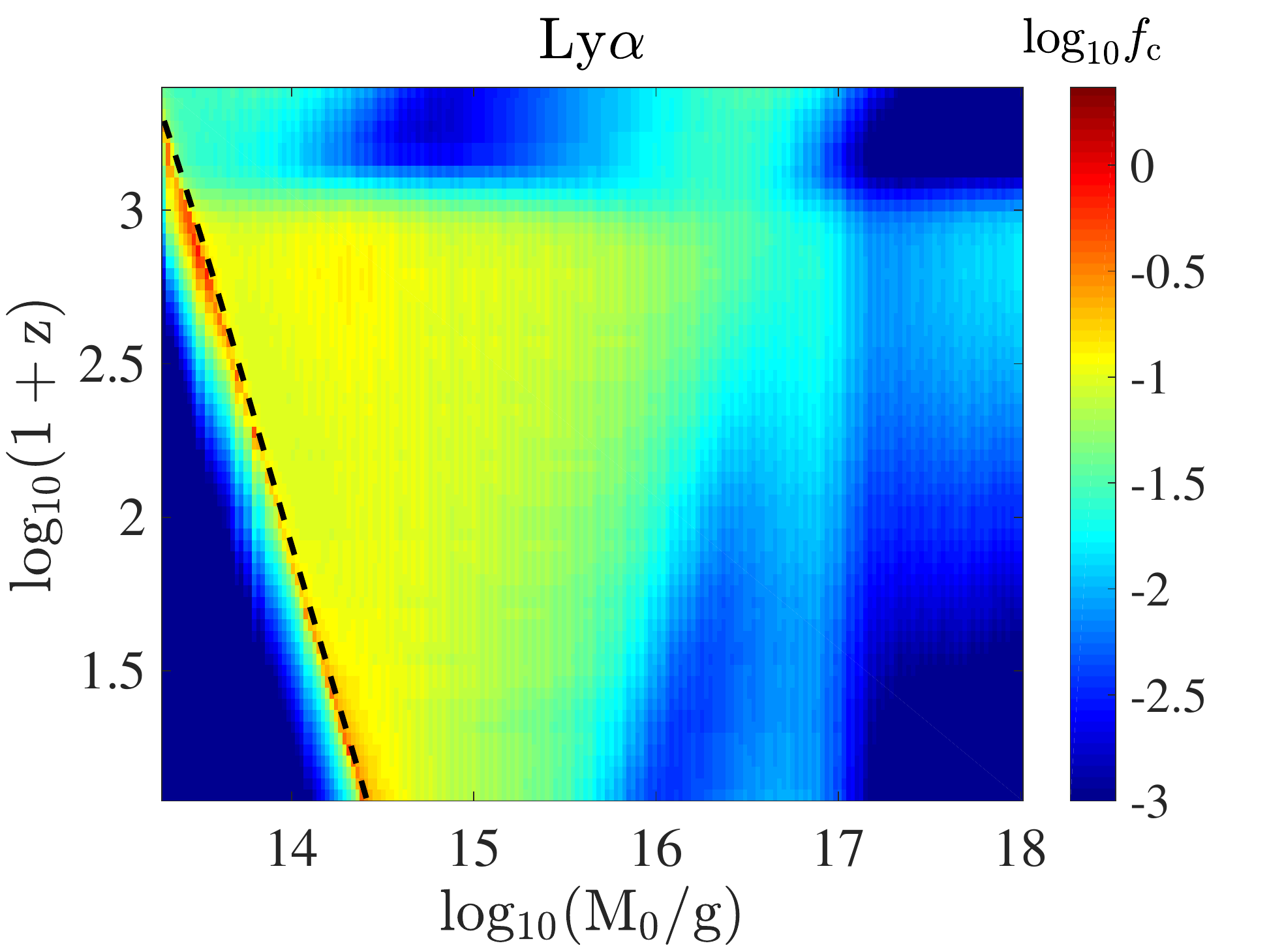}  \includegraphics[width=5.2cm]{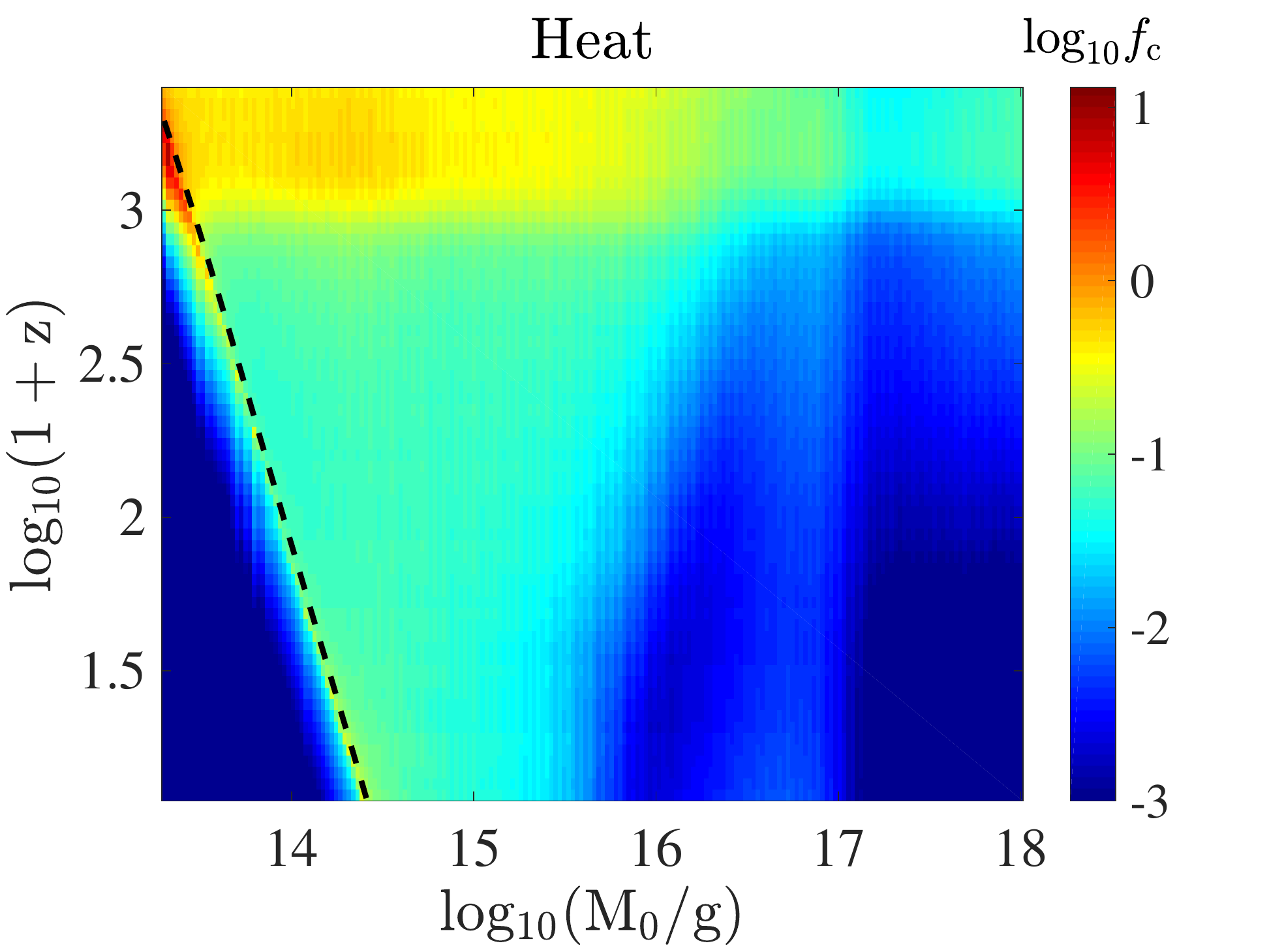}}
\subfigure{\includegraphics[width=5.2cm]{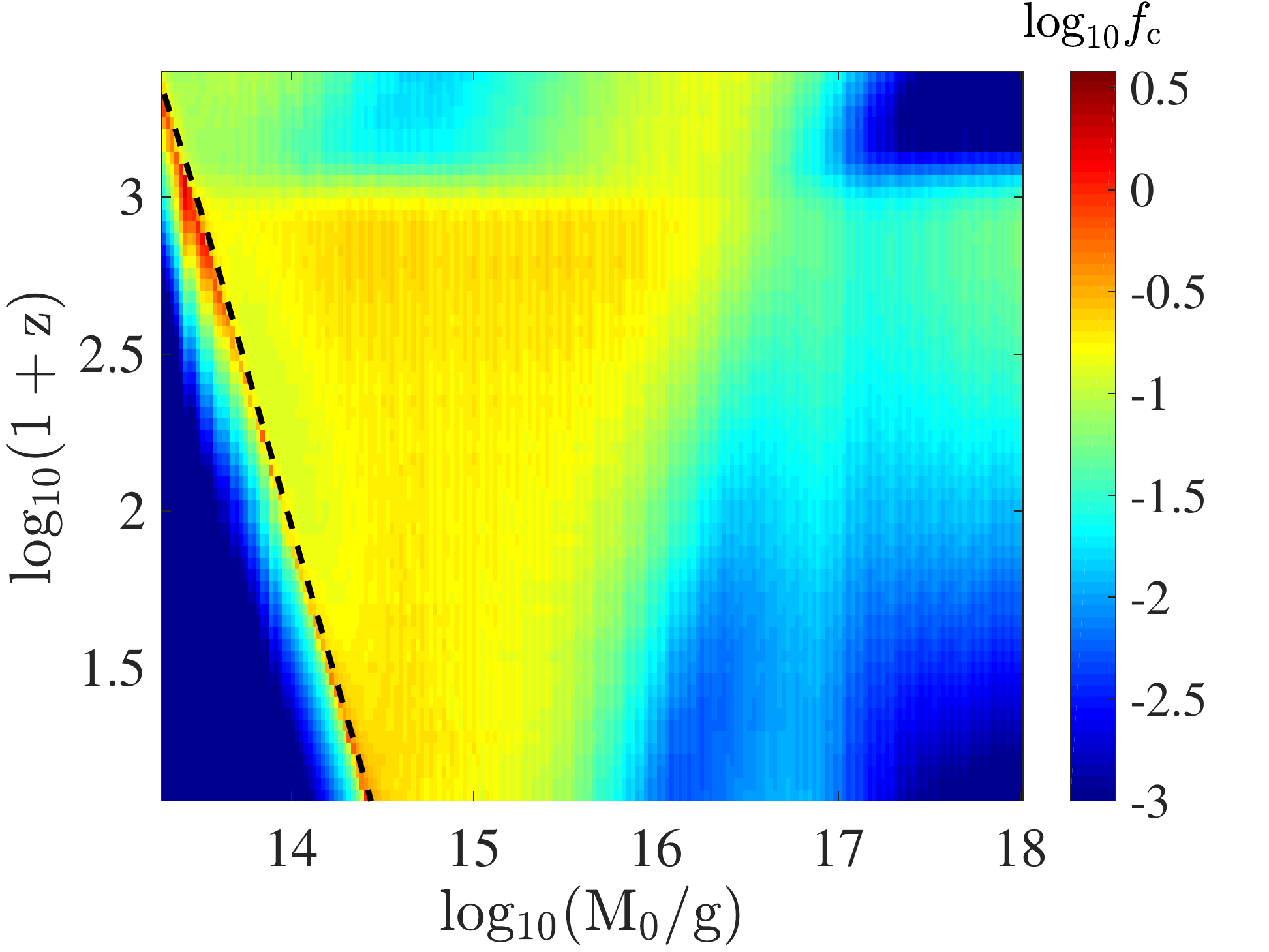} \includegraphics[width=5.2cm]{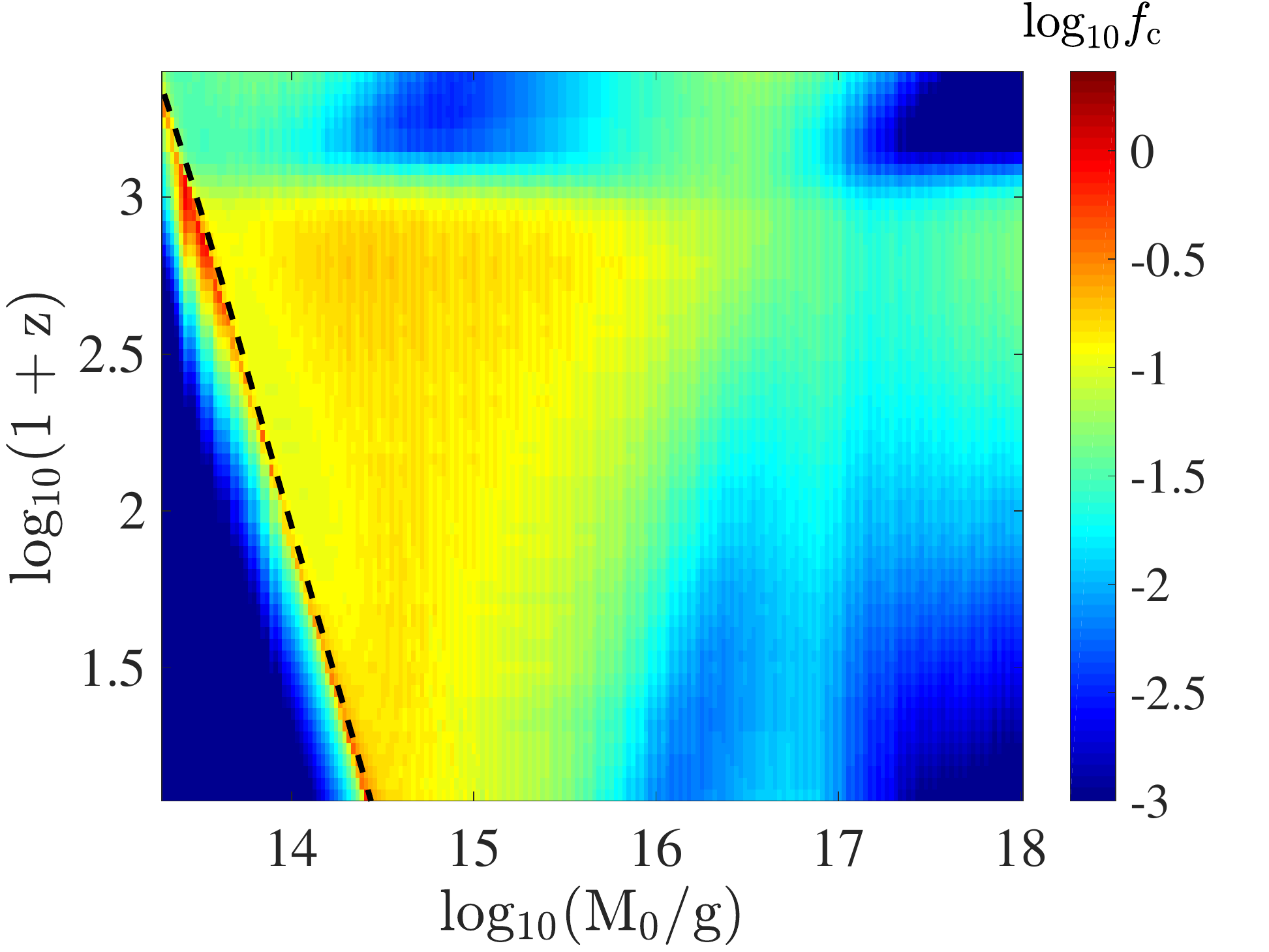}  \includegraphics[width=5.2cm]{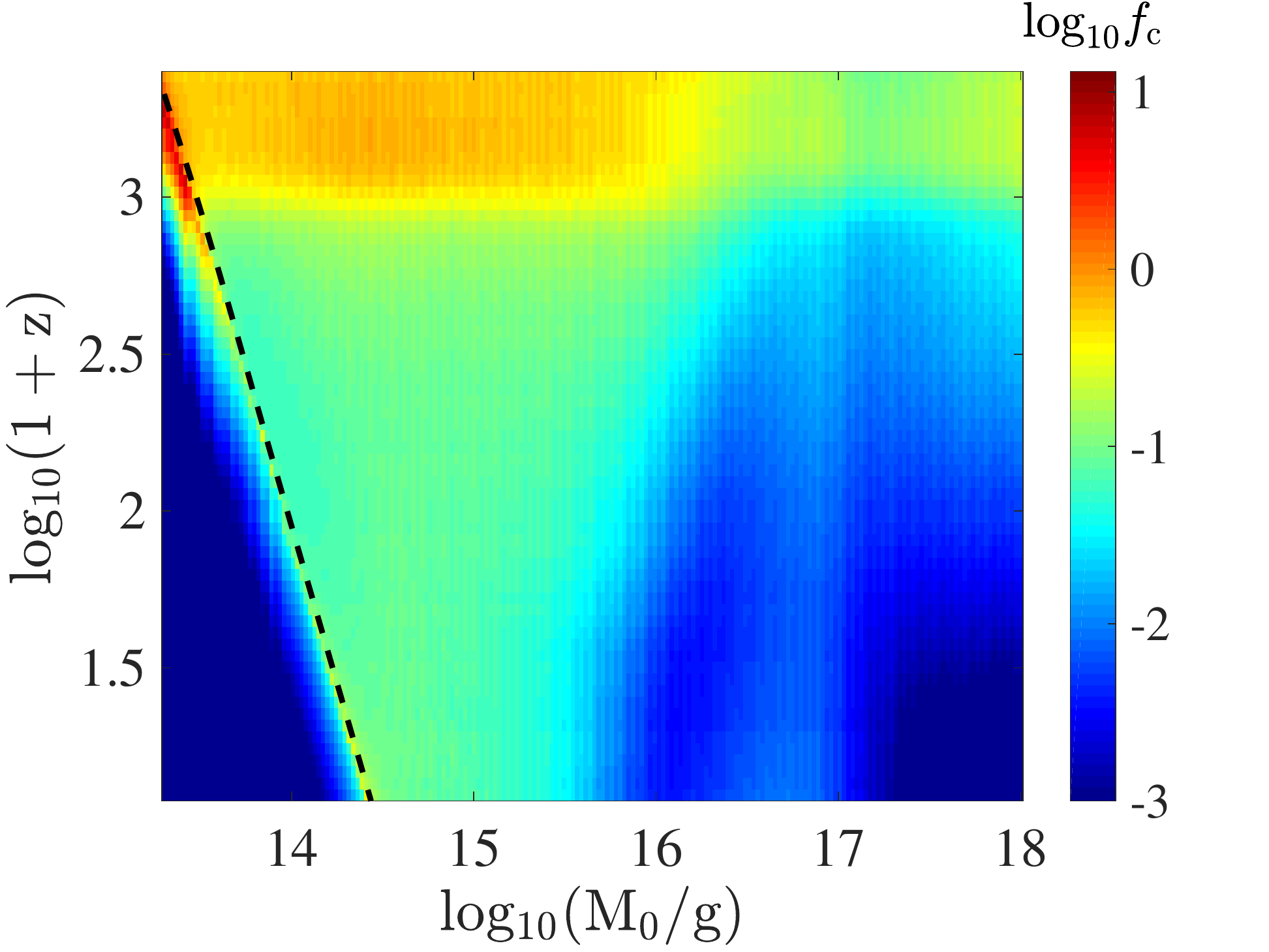}}
\subfigure{\includegraphics[width=5.2cm]{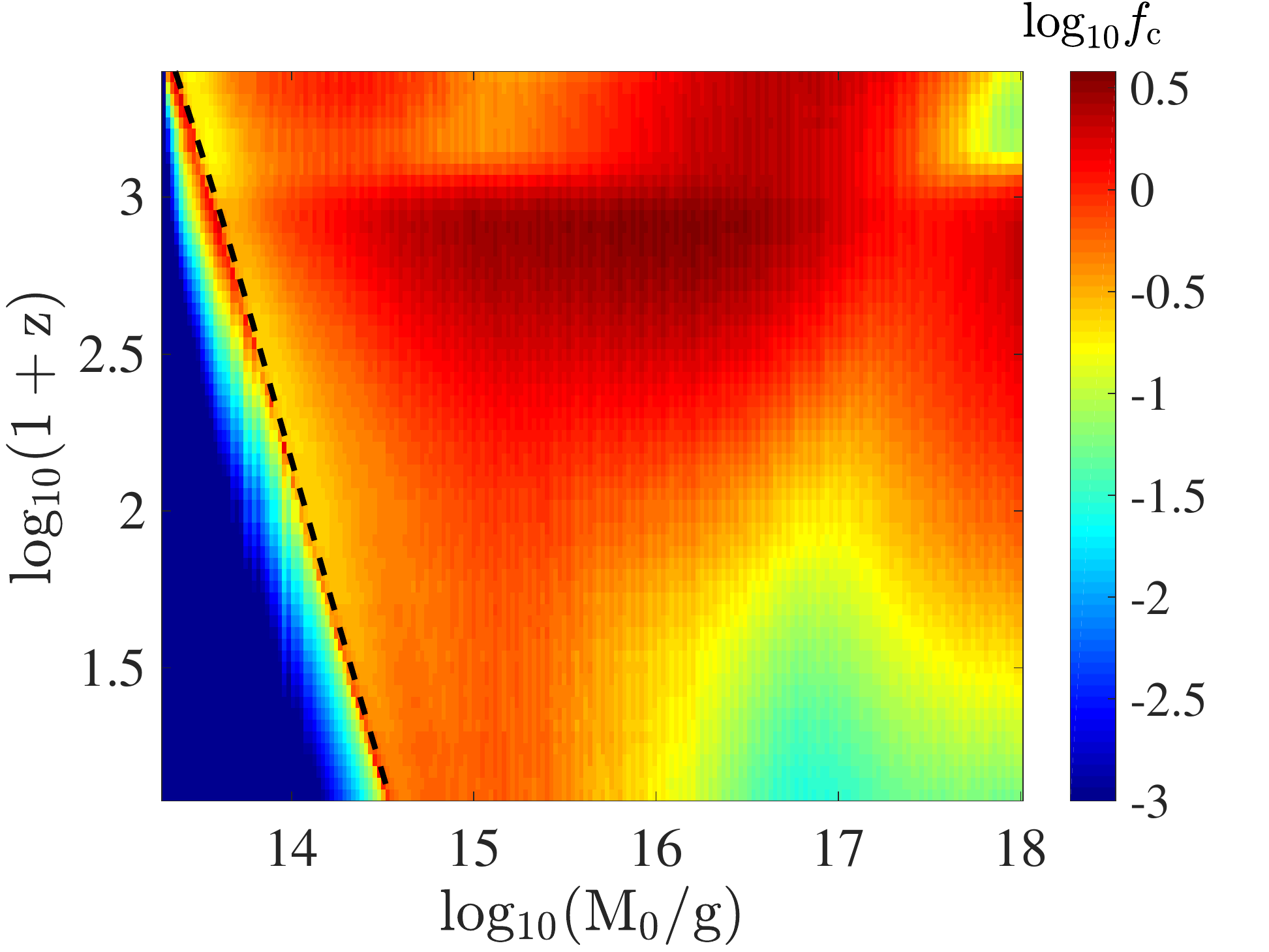} \includegraphics[width=5.2cm]{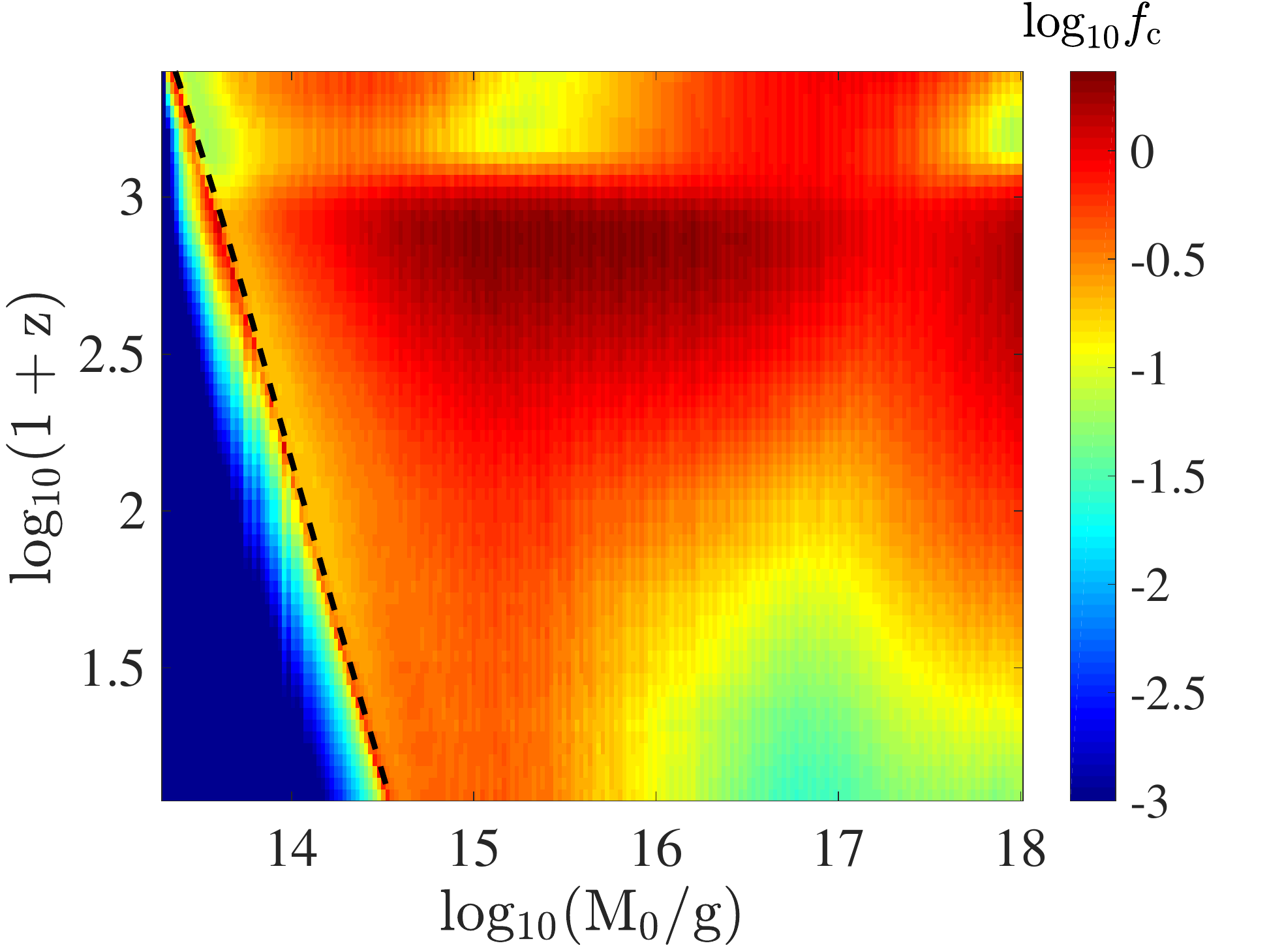}  \includegraphics[width=5.2cm]{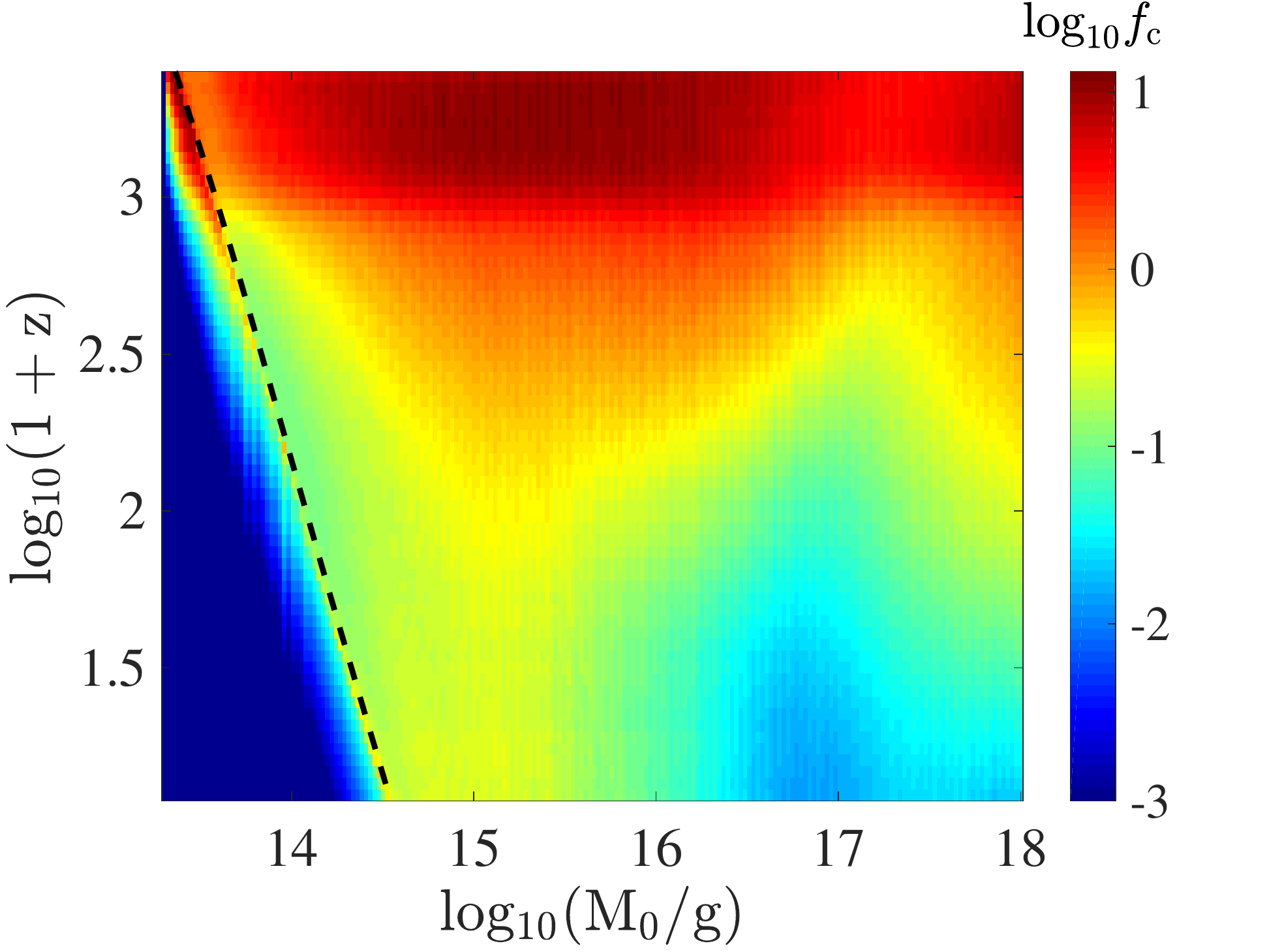}}
\caption{
Effective deposition efficiency $f_{\r{c}}$ defined in Eq. (\ref{kdsfkj887}) for monochromatic PBHs.
The deposition channels are hydrogen ionization (left), hydrogen excitation (middle) and heating (right),
the top, middle and bottom rows correspond to $a_0$ values of 0, 0.5 and 0.999 respectively.
Redshifts when the PBH reaches its' end of life are indicated by black dashed lines.
$f_\r{c}$ values lower than $10^{-3}$ have been set to $10^{-3}$ here for illustrative purpose (dark blue regions).
}
\label{EFF_Plot}
\end{figure*}
$f_\r{c}$ can be pre-computed numerically,
such that deposition rate at any given redshift can be recovered through,
\be
\label{k887}
\left[
\frac{{\rm{d}}E}
{{\rm{d}}V{\rm{d}}t}
\right]_{\rm{dep,c}}^{\delta}
(z)
=
f_{\rm{c}}
(z)
\times
\left[
\frac{{\rm{d}}E}
{{\rm{d}}V{\rm{d}}t}
\right]_{\rm{inj}}'.
\ee

Using Eqs. (\ref{Dep_EFF_EQds},\ref{888eygswdgvywy},\ref{kdsfkj887}),
after re-arranging terms,
$f_{\r{c}}$ reads
\be
\begin{aligned}
f_{\r{c}}
(M_0,a_0,z)
&=
\frac{H(z)}{5.34 \times 10^{25}}
\left(\frac{M_0}{\r{g}}\right)^2
\int
\frac
{\r{d} z'}
{(1+z') H(z')}
\\
&
\times
\sum_{\alpha = \gamma, e^{\pm}}
\int \r{d}E
\ 
E
\mathcal{T}^{\alpha}_{\r{c}}(z,E,z')
\left[
\frac{\r{d}N^{\alpha}}{\r{d}E\r{d}t}
\right]'
\Theta(\tau_{\r{PBH}} - t')
\ 
\r{s \cdot g^{-1}}
.
\end{aligned}
\label{fdgsfehfh}
\ee
Fig.~\ref{EFF_Plot} shows $f_{\r{c}}$ for all deposition channels and $a_0$ values considered in this work.
Since one must perform integration in Eqs. (\ref{Dep_EFF_EQ}) and (\ref{fdgsfehfh}) over all injection redshifts prior to $z$,
$f_\r{c}$ at any given redshift receives contribution from particle injected at earlier redshifts,
therefore deposition rate at a certain redshift can be nonzero even when the injection source, i.e PBH, has already vanished.
Another interesting feature about Fig.~\ref{EFF_Plot} is the sudden increase in $f_{\r{c}}$ when a PBH reaches the end of its lifetime,
caused by the 
surge of energy injection rate during PBH explosion.

\subsection{Extended PBH distribution}
In analogy to Eq. (\ref{dsfghw232}),
for PBHs with extended distribution,
the differential form of the particle injection history $\mathcal{I^{\alpha}}$ is given by
\be
\r{d}
\mathcal{I}^{\alpha}
=
\left[
\frac{\r{d}N^{\alpha}}{\r{d}E\r{d}t}
\right]
\left(M\right)
\r{d} n_{\r{PBH}}
\label{hubuy76}
\ee
here $\r{d} n_{\r{PBH}}$ is the number density of PBHs in mass range $[M, M+\r{d}M]$,
\be
\begin{aligned}
\r{d}
n_{\r{PBH}}
& =
(1+z)^3
\frac{\r{d}\rho}{M}
\\
&=
f_{\r{PBH}}
\Omega_{\r{c}}
\rho_{\r{cr}}
(1+z)^3
\Psi(M,t)
\frac{\r{d} M}{M}
\\
&=
f_{\r{PBH}}
\Omega_{\r{c}}
\rho_{\r{cr}}
(1+z)^3
\Psi(M_0,t_0)
\frac{\r{d} M_0}{M_0}
\end{aligned}
\ee
where we used Eqs. (\ref{shffw67tyh},\ref{dsfefhjedwjk_2}) in the second line and Eq. (\ref{dksgdshj_1}) in the third line.
Inserting this equation into Eq. (\ref{hubuy76}) and integrating over all survived $M_0$ values gives,
\be
\mathcal{I}^{\alpha}
(z)
=
f_{\r{PBH}}
\Omega_{\r{c}}
\rho_{\r{cr}}
(1+z)^3
\int_{M^{\r{min}}_{0}(z)}^{\infty}
\frac{\r{d}M_0}{M_0}
\left[
\frac{\r{d}N^{\alpha}}{\r{d}E\r{d}t}
\right]
(M)
\Psi(M_0,t_0)
\label{xsnif938r7y}
\ee
$M^{\r{min}}_{0}(z)$ is the mass of PBH whose lifetime $\tau_{\r{PBH}}$ equals $t(z)$,
or equivalently the smallest initial PBH mass that can survive to redshift $z$.
Comparing this equation with Eq. (\ref{dsfghw232}), one finds that
\be
\mathcal{I}^{\alpha}
(z)
=
\int_0^{\infty}
\r{d}M_0
\ 
\mathcal{I}^{\alpha, \delta}
(M_0,z)
\Psi(M_0,t_0)
.
\label{dsfjhejbvdjy675ty}
\ee
We have moved lower limit of $M_0$ integration from $M^{\r{min}}_{0}(z)$ to $0$ because 
$\mathcal{I}^{\alpha, \delta}(M_0,z)$ given by Eq. (\ref{dsfghw232}) automatically vanishes once $M_0$ reaches its end of life,
therefore $M_0$ integration from $0$ to $M^{\r{min}}_{0}(z)$ does not change the results of Eq. (\ref{dsfjhejbvdjy675ty}).
Inserting Eq. (\ref{dsfjhejbvdjy675ty}) into Eq. (\ref{Dep_EFF_EQ}) gives the deposition rate for extended PBH distribution,
\be
\begin{aligned}
\left[
\frac{{\rm{d}}E}
{{\rm{d}}V{\rm{d}}t}
\right]_{\rm{dep,c}}
(z)
& =
(1+z)^3 H(z)
\int
\frac
{\r{d} z'}
{(1+z')^4 H(z')}
\\
&
\times
\sum_{\alpha = \gamma, e^{\pm}}
\int \r{d}E
\ 
E
\mathcal{T}^{\alpha}_{\r{c}}(z,E,z')
\left[
\int_0^{\infty}
\r{d}M_0
\ 
\mathcal{I}^{\alpha, \delta}
(M_0,z')
\Psi(M_0,t_0)
\right]
\\
&=
\int_0^{\infty}
\r{d}M_0
\ 
\Psi(M_0,t_0)
(1+z)^3 H(z)
\\
&
\times
\int
\frac
{\r{d} z'}
{(1+z')^4 H(z')}
\sum_{\alpha = \gamma, e^{\pm}}
\left[
\int \r{d}E
\ 
E
\mathcal{T}^{\alpha}_{\r{c}}(z,E,z')
\mathcal{I}^{\alpha,\delta}(E,z')
\right]
,
\end{aligned}
\label{Dep_EFF_EQgugfhisdh}
\ee
which can be simplified as an integration over deposition rate for a monochromatic distribution at $M_0$,
\be
\begin{aligned}
\left[
\frac{{\rm{d}}E}
{{\rm{d}}V{\rm{d}}t}
\right]_{\rm{dep,c}}
(z)
=
\int_{2 \times 10^{13} \r{g}}^{10^{18}\r{g}}
\r{d}M_0
\ 
\Psi(M_0,t_0)
\left[
\frac{{\rm{d}}E}
{{\rm{d}}V{\rm{d}}t}
\right]^{\delta}_{\rm{dep,c}}(M_0,z)
.
\end{aligned}
\label{dsjghfcvfdbjz8765}
\ee
where the integrated $M_0$ range has been set to the mass window considered in this work.
Similarly one can also show that injection rate for extended distribution follows an analogous relation,
\be
\begin{aligned}
\left[
\frac{{\rm{d}}E}
{{\rm{d}}V{\rm{d}}t}
\right]_{\rm{inj}}
(z)
=
\int_0^{\infty}
\r{d}M_0
\ 
\Psi(M_0,t_0)
\left[
\frac{{\rm{d}}E}
{{\rm{d}}V{\rm{d}}t}
\right]^{\delta}_{\rm{inj}}(M_0,z)
.
\end{aligned}
\label{dsjghfcvfdbjINJ}
\ee
For typical extended mass distributions, we will consider log-normal, critical collapse and power-law scenarios as in Eqs.~\ref{fdsdsahfg54esdf1qw}-\ref{MdsgfF_3}. The IGM temparture rise and $T_{21}$ correction at early reionization epoch are numerically evaluated and compared to experimentally measured upper limits. Corresponding limits on the PBH injection rate are given in the next section.

\section{21-cm limits}
\label{sdgsfhgvhdhh45rtgh}
IGM temperature and ionization evolutions, after accounting for PBH heating/ionization effects as in Eqs. (\ref{dsfeffbshfdhvb761},\ref{dsfeffbshfdhvb76}), are numerically computed by
modified {\tt HyRec} package~\cite{Ali-Haimoud:2010hou}. IGM heating's impact on 21-cm signal reveals after Wouthuysen-Field effect takes place, where the hydrogen spin temperature $T_{\rm S}$ becomes bound to gas temperature $T_{\rm K}$.
PBH heating increases gas temperature $T_{\r{K}}$ which is typically closely coupled to $T_{\r{S}}$ at redshift $z=17$~\cite{Pritchard:2011xb,Mesinger:2010ne},
and in turn damps the amplitude of $T_{21}$ signal observed by EDGES.
Here we derive our 21-cm upper bounds on $f_{\r{PBH}}$ by imposing that PBH does not over-heats the IGM gas, or equivalently raises $T_{21}$ beyond $-150\ \r{mK}$,
\be
T_{21}
<
-150
\r{mK},
\ 
{\rm at\ }z=17.
\label{cnvbasx234567ijhgfddtvh}
\ee
This choice is 
noticeably 
higher than both the EDGES 99\% C.L. upper limit of $-300\ \r{mK}$,
and the standard $\Lambda \r{CDM}$ lower limit of $-210\ \r{mK}$~\cite{Barkana:2018lgd,Cheung:2018vww}.
Generally $T_{\r{S}}$ is coupled to both $T_{\r{CMB}}$ and $T_{\r{K}}$ through collisional coupling and Wouthuysen-Field effect~\cite{Pritchard:2011xb,1958PIRE...46..240F,Mittal:2020kjs},
such that at any redshifts one can expect either $T_{\r{K}} \le T_{\r{S}} \le T_{\r{CMB}}$ or $T_{\r{CMB}} \le T_{\r{S}} \le T_{\r{K}}$.
Since $T_{21}$ at $z=17$ has been measured to be in absorption ($T_{21}<0$) by EDGES, 
one can infer that at this redshift $T_{\r{K}} \le T_{\r{S}} < T_{\r{CMB}}$,
combining this with Eqs. (\ref{fdgfhejwuy7},\ref{cnvbasx234567ijhgfddtvh}) and $T_{\r{CMB}}= 2.73 \ (1+z) \ \r{K}$,
we arrive at the final equation for solving our 21-cm limit on $f_{\r{PBH}}$,
\be
T_{\r{K}}
<
49.1
\left[
1+\frac{4.14}{
1-x_{\r{e}}
}
\right]^{-1}
\ 
\r{K}
,
\ 
z=17.
\label{dsdghwi6754erdfg}
\ee
where we have set $x_{\r{HI}} = 1-x_{\r{e}}$ because the effect of helium reionization is negligible at this redshift~\cite{Puchwein:2018arm,Sokasian:2001xh},
such that $n_{\r{e}} \simeq n_{\r{p}}$ and $x_{\r{HI}} \equiv n_{\r{HI}} / ( n_{\r{p}}+n_{\r{HI}}) \simeq 1-x_{\r{e}}$.

Note that PBHs can also leave their footprints on CMB anisotropy mainly by raising the ionization level $x_{\r{e}}$ which affects the propagation of CMB photons.
On CMB temperature and polarization anisotropy spectrum,
this can suppress small scale correlations and shift polarization peak locations~\cite{Padmanabhan:2005es}.
As a comparison with 21-cm results,
we also show $f_{\r{PBH}}$ limit from the CMB.
We interfaced the {\tt CAMB} codes~\cite{Lewis:1999bs} with our modified {\tt HyRec} for calculation of CMB anisotropy spectra in presence of Hawking radiation,
and our CMB constraints on $f_{\r{PBH}}$ are given by MCMC analysis of {\it{Planck}} 2018 data~\cite{Planck:2019nip} using the {\tt CosmoMC} package~\cite{Lewis:2002ah,Lewis:2013hha}.
Specifically, 
the datasets used are the high-$\ell$ plik-lite TTTEEE likelihood, 
low-$\ell$ TT and EE likelihood and the lensing likelihood.
In addition to our PBH parameters,
all six base $\Lambda {\r{CDM}}$ parameters are also varied during MCMC analysis.

\begin{figure*}[tp]
\centering
\subfigbottomskip=-200pt
\subfigcapskip=-7pt
\subfigure{\includegraphics[width=8cm]{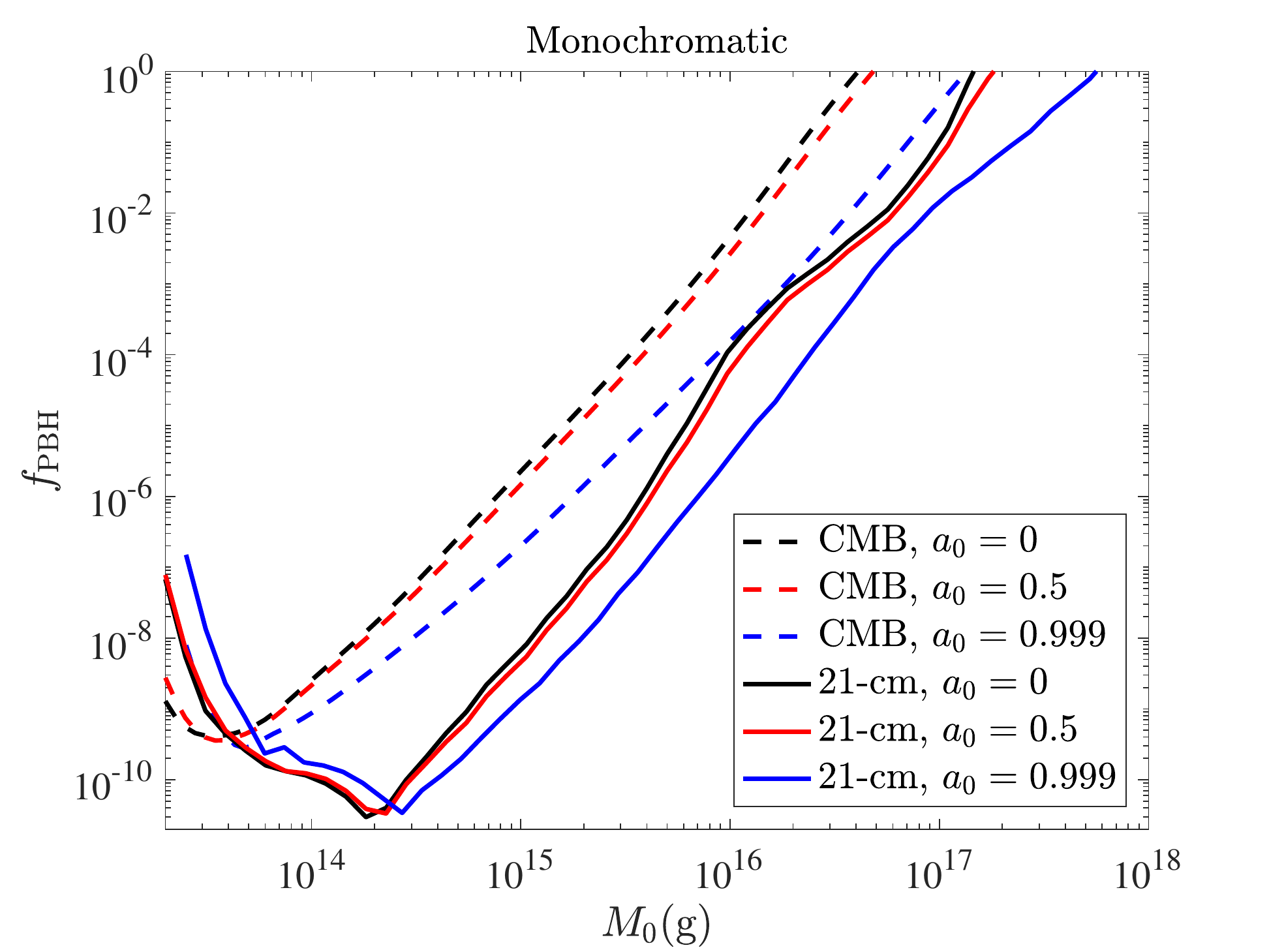}  \includegraphics[width=8cm]{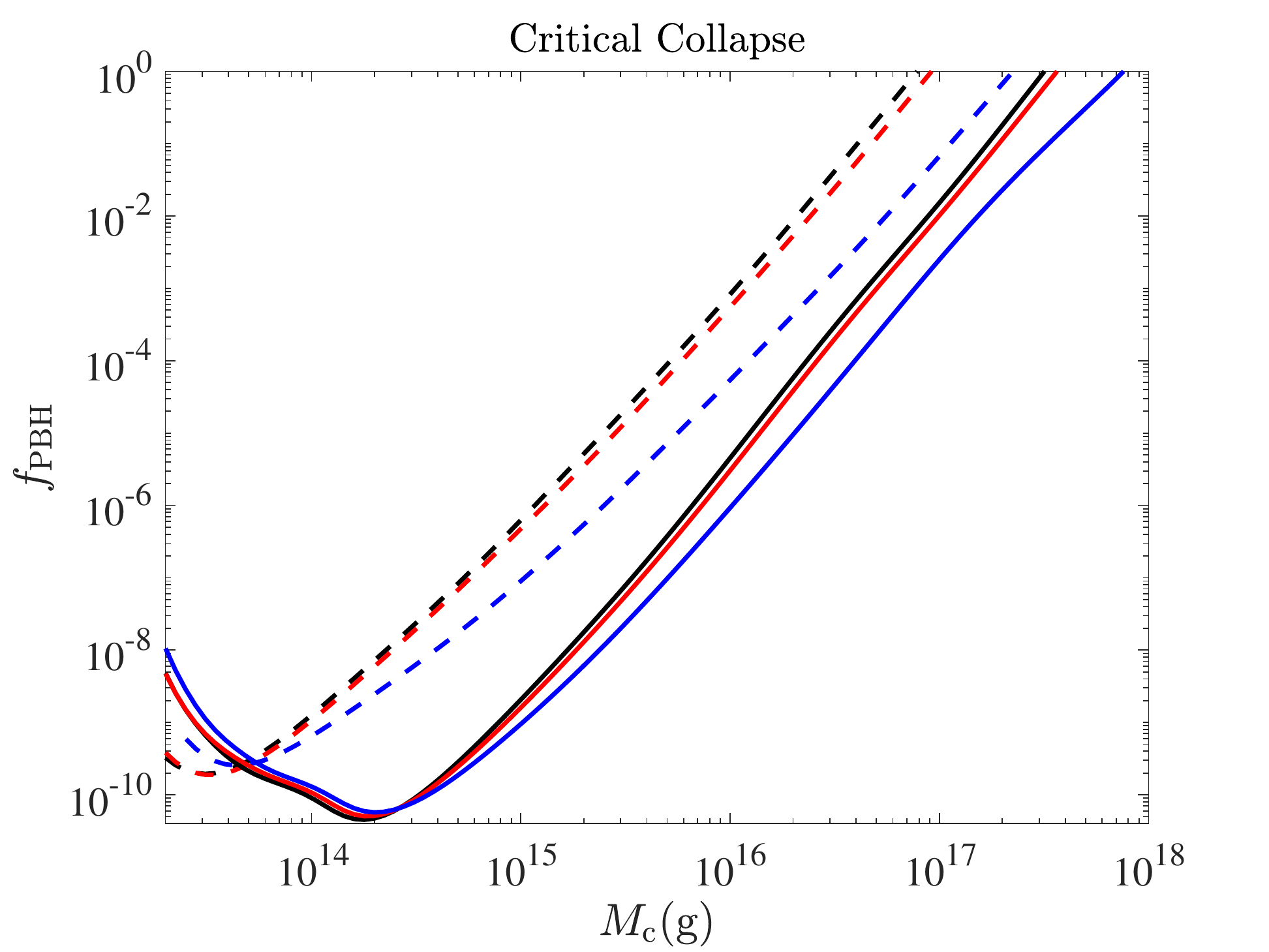}}
\caption{
$f_{\rm{PBH}}$ upper bounds for monochromatic (left) and Critical Collapse (right) distributions.
Constraints from 21-cm are shown in solid lines,
dashed lines show marginalized 95\% C.L. limits given by MCMC analysis of CMB anisotropy data from {\it{Planck}}.
Black, red and blue curves show constraints for initial Kerr spin values of 0, 0.5 and 0.999 respectively.
The legend applies to both panels.
}
\label{Re_1D}
\end{figure*}

\begin{figure*}[tp] 
\centering
\subfigbottomskip=0pt
\subfigcapskip=0pt
\subfigure[{$a_0=0$, Schwarzchild}]{\includegraphics[width=7.8cm]{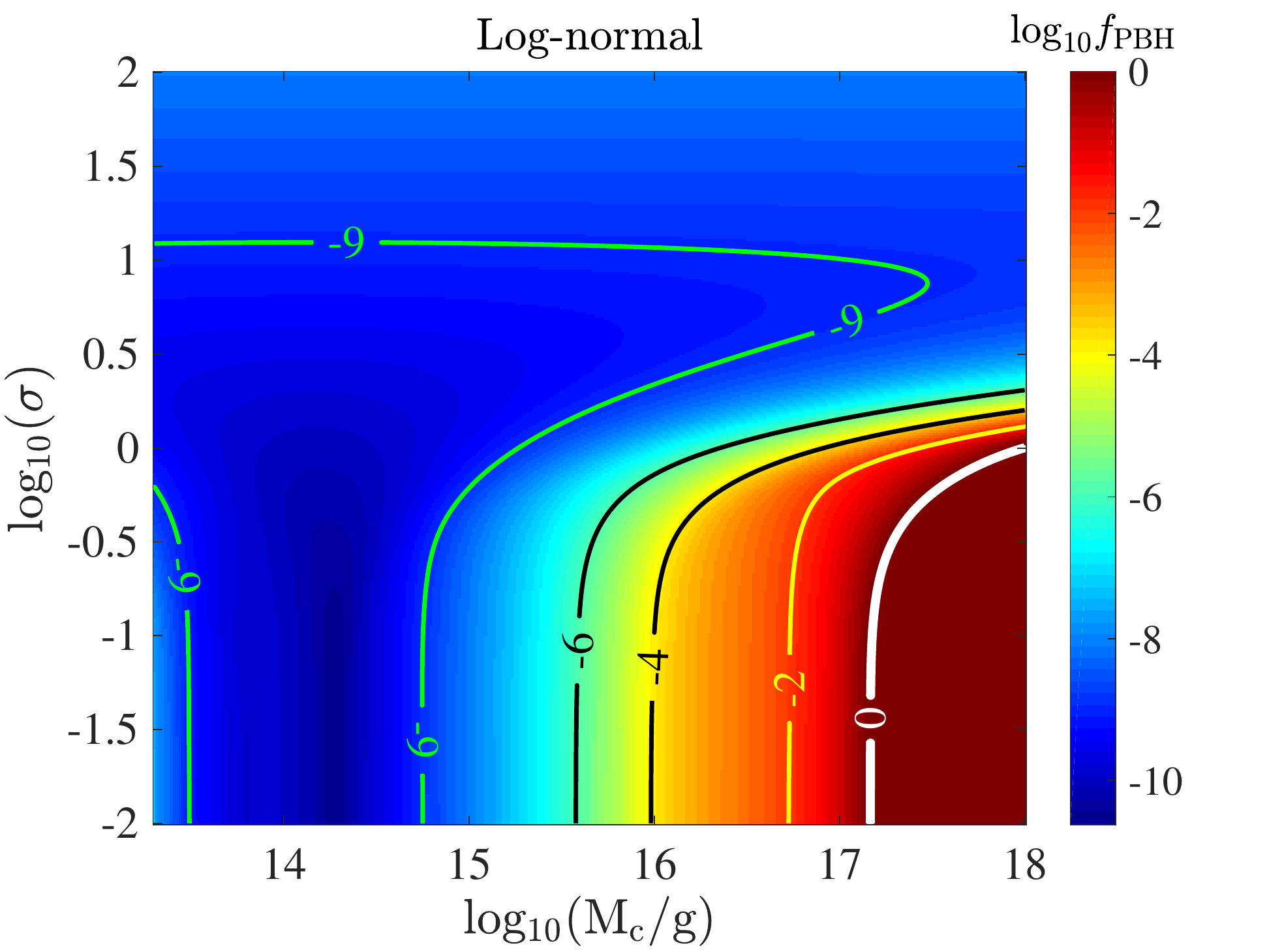}  \includegraphics[width=7.8cm]{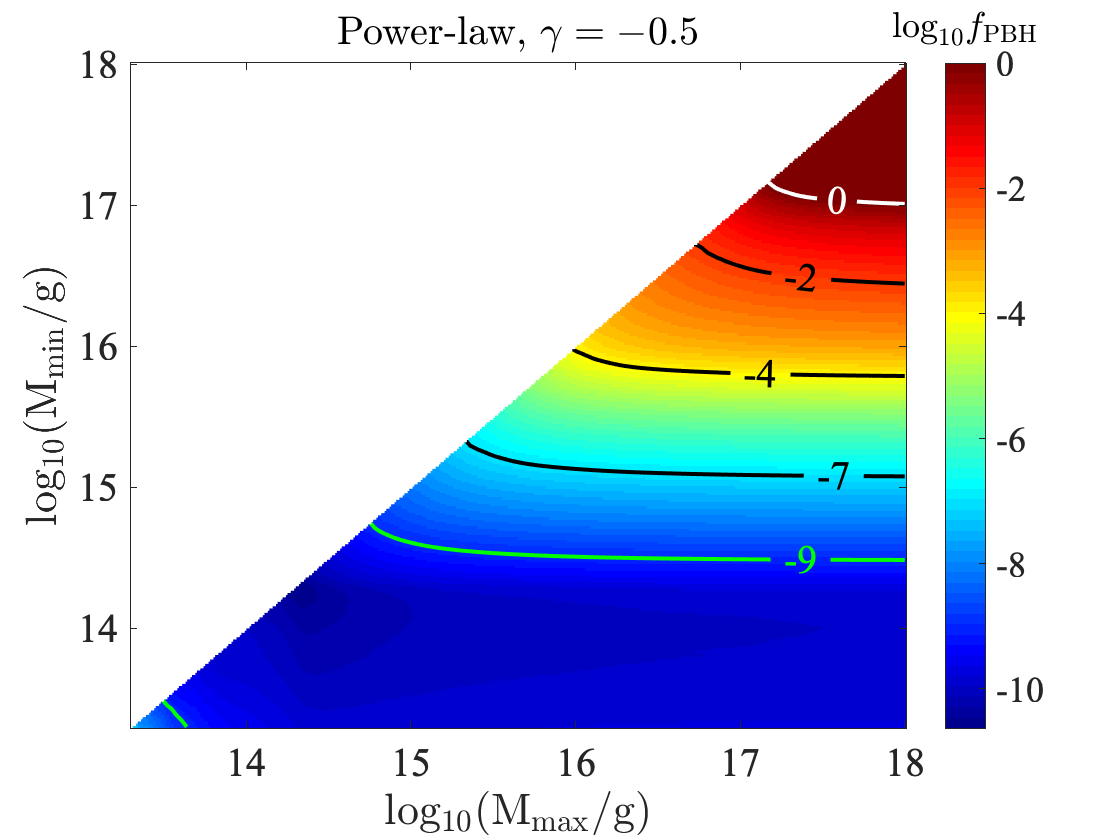}}
\subfigure[{$a_0=0.5$}]{\includegraphics[width=7.8cm]{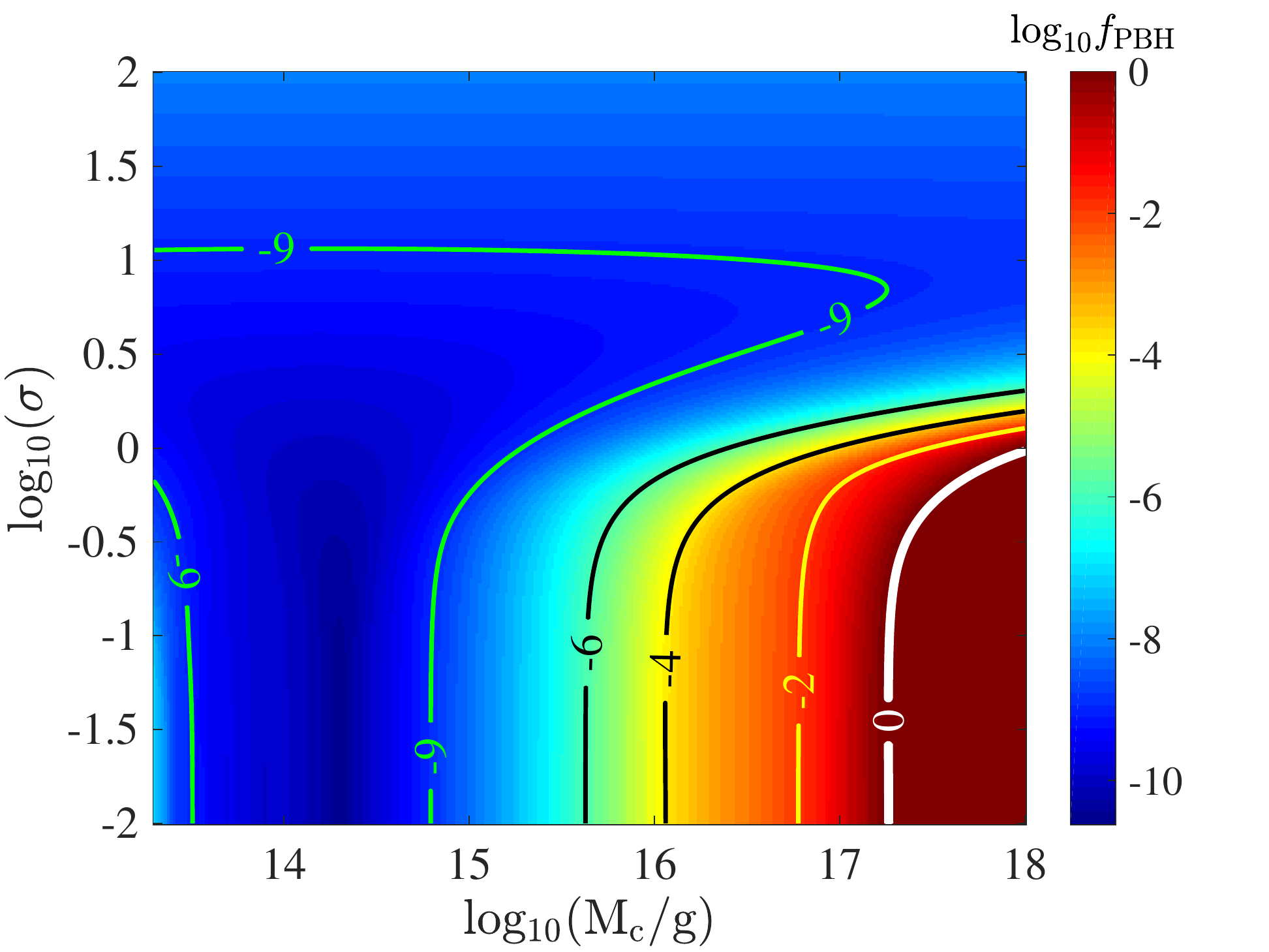}  \includegraphics[width=7.8cm]{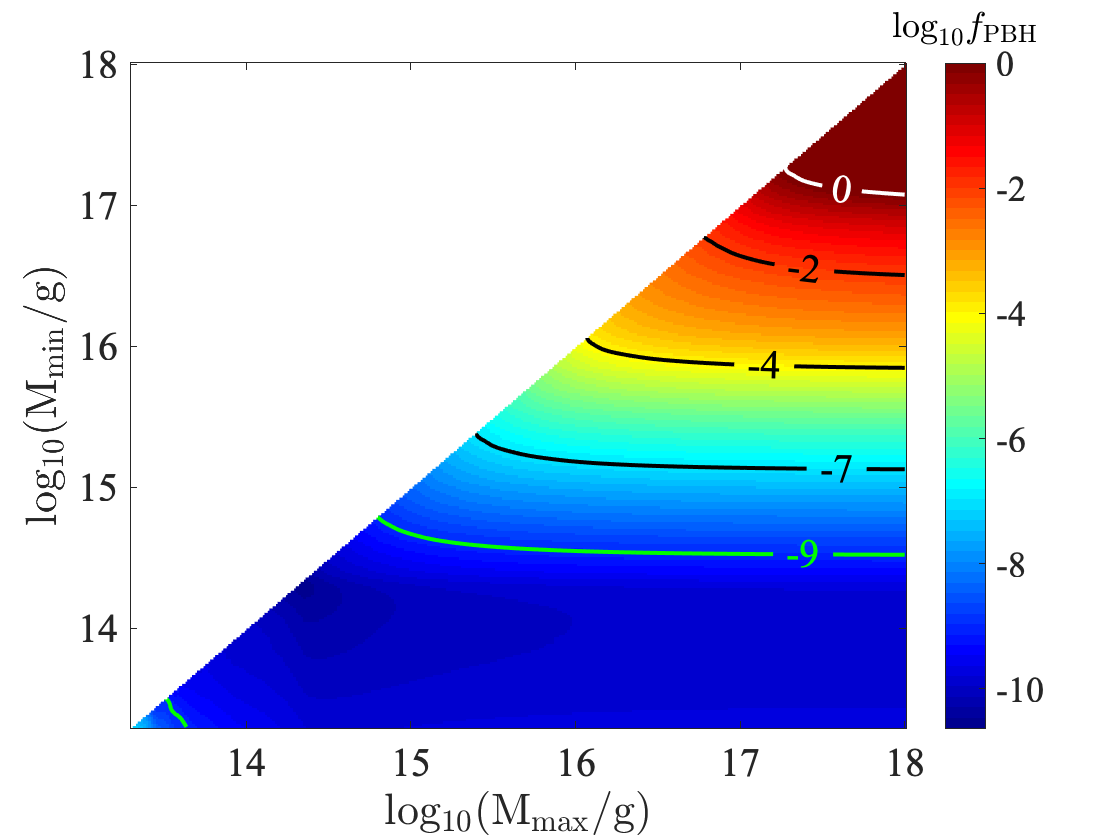}}
\subfigure[{$a_0=0.999$}]{\includegraphics[width=7.8cm]{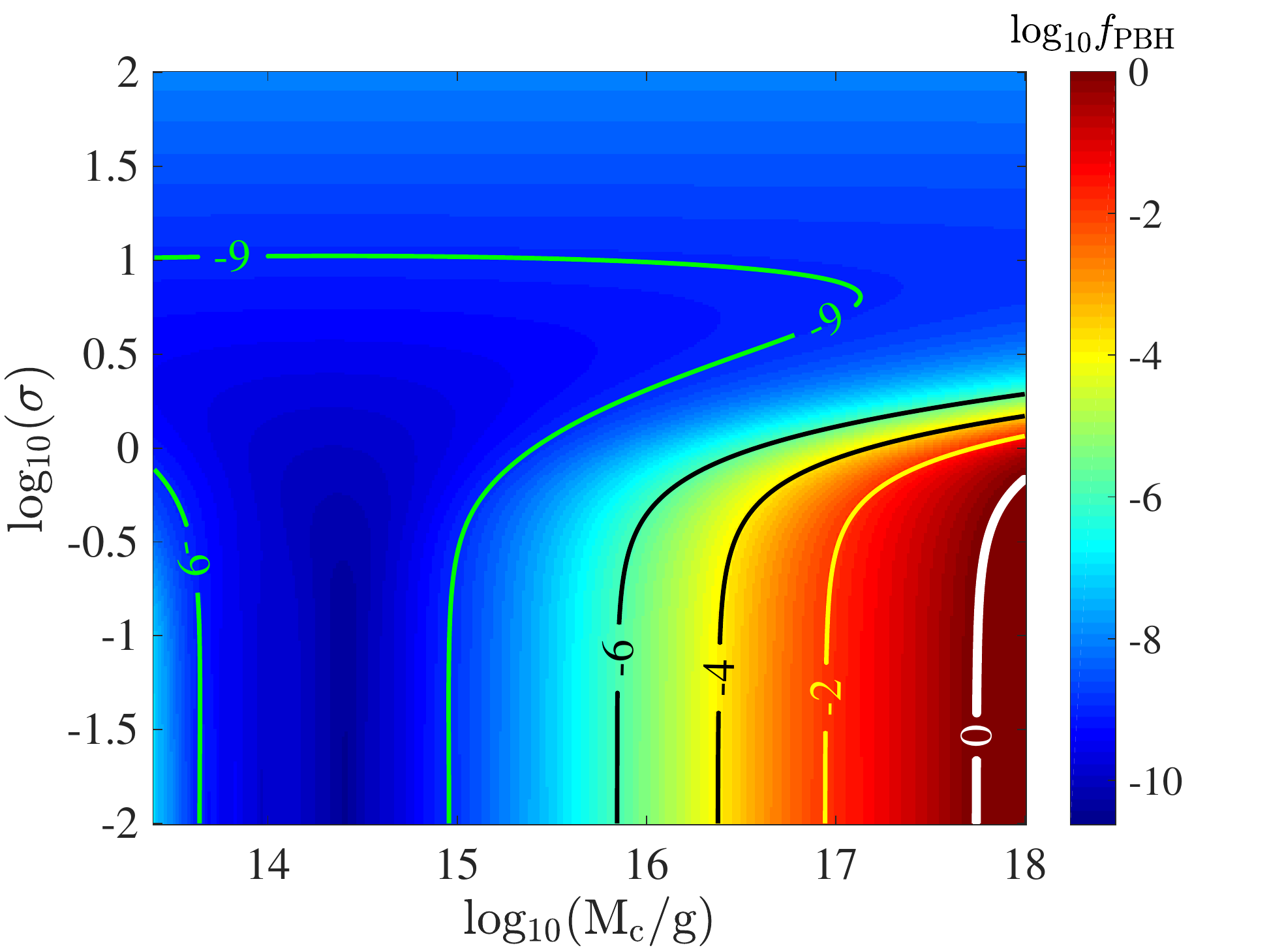}  \includegraphics[width=7.8cm]{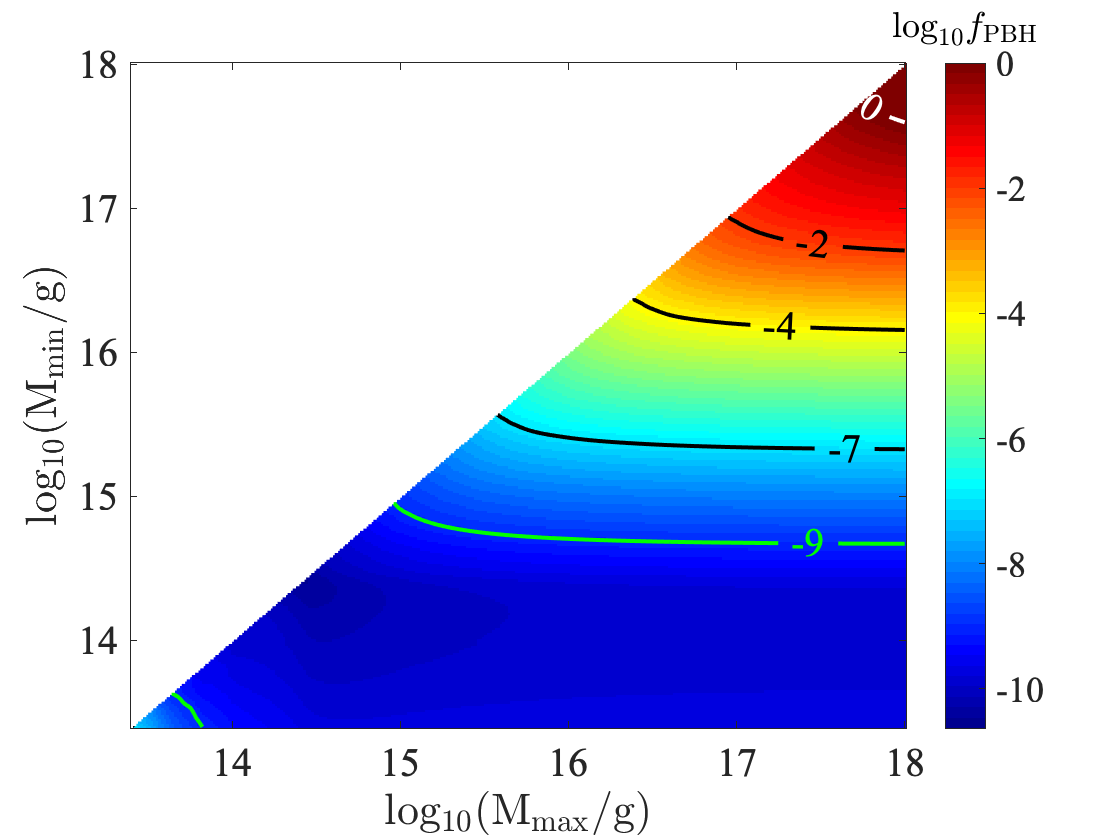}}
\caption{
21-cm upper bounds on $f_{\rm{PBH}}$ for log-normal model (left) and a power-law (right) distribution with $\gamma=-0.5$,
corresponding to PBHs formed in radiation dominated epoch.
The top, middle and bottom panels correspond to initial spins of $a_0=0$, $0.5$ and $0.999$ respectively.
White contours show regions in which PBH can account for all DM ($f_{\r{PBH}} =1$).
}
\label{Re_2D}
\end{figure*}

\begin{figure*}[tp] 
\centering
\subfigbottomskip=0pt
\subfigcapskip=0pt
\subfigure[{$a_0=0$, Schwarzchild}]{\includegraphics[width=7.8cm]{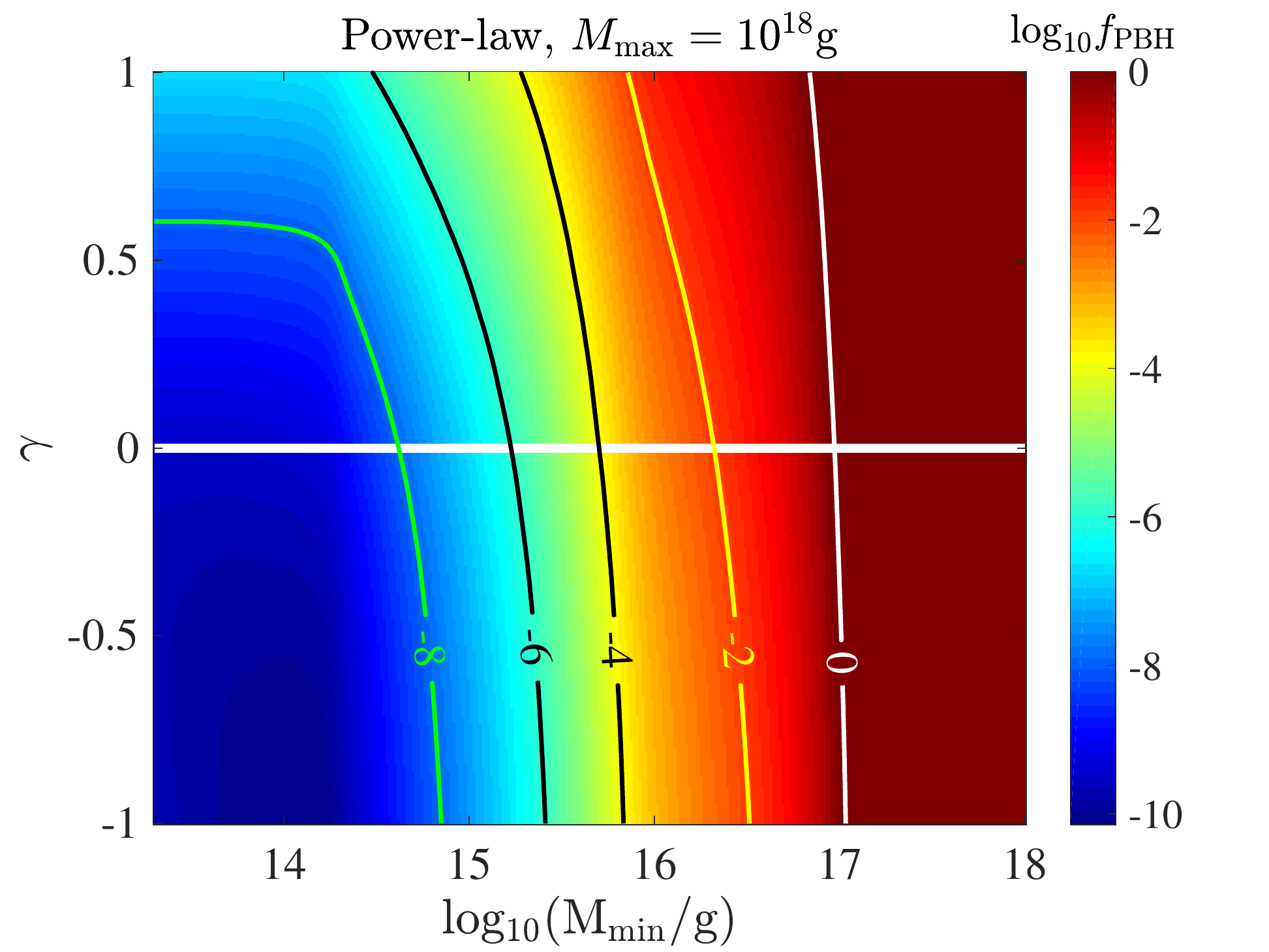}  \includegraphics[width=7.8cm]{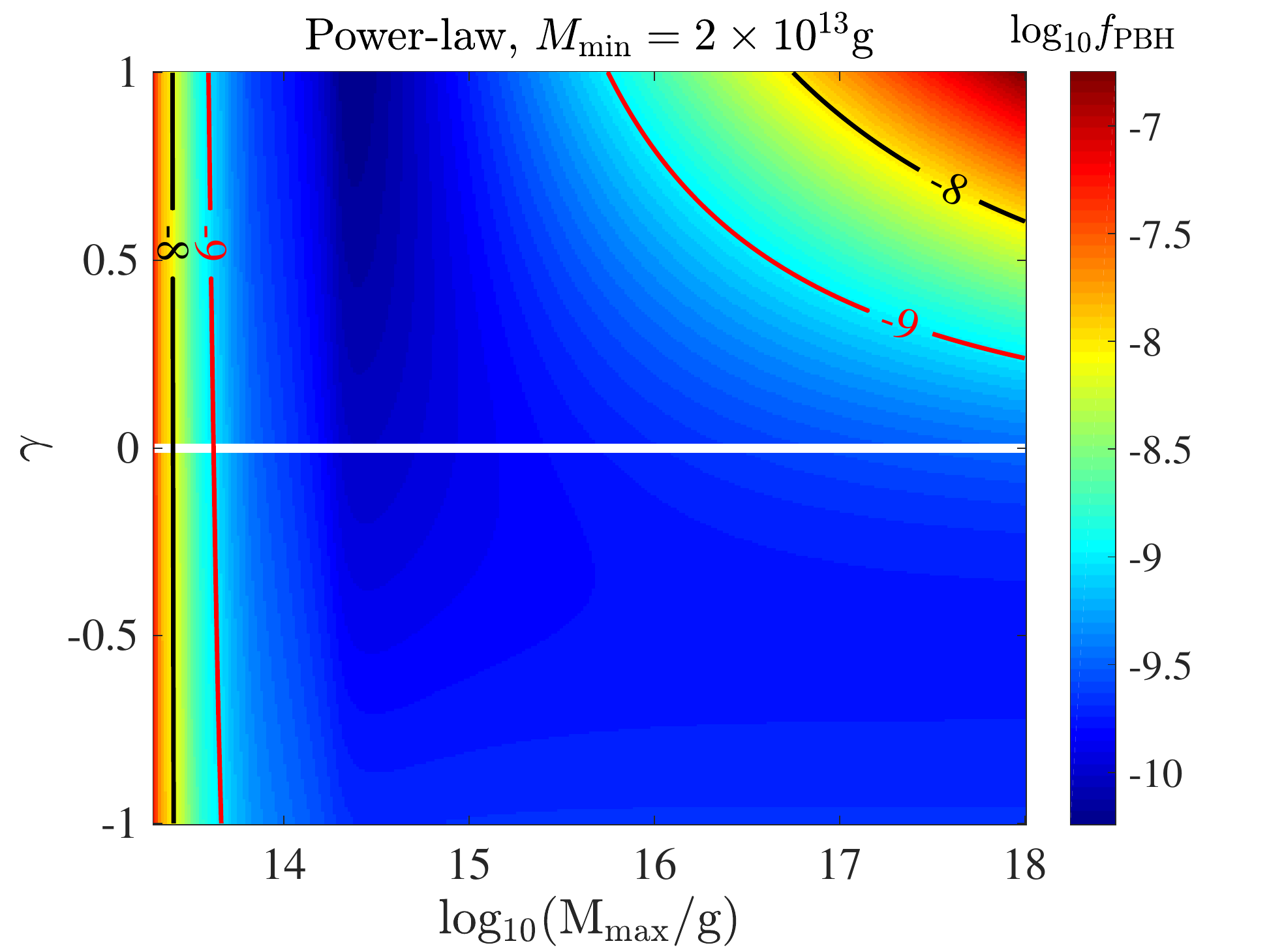}}
\subfigure[{$a_0=0.5$}]{\includegraphics[width=7.8cm]{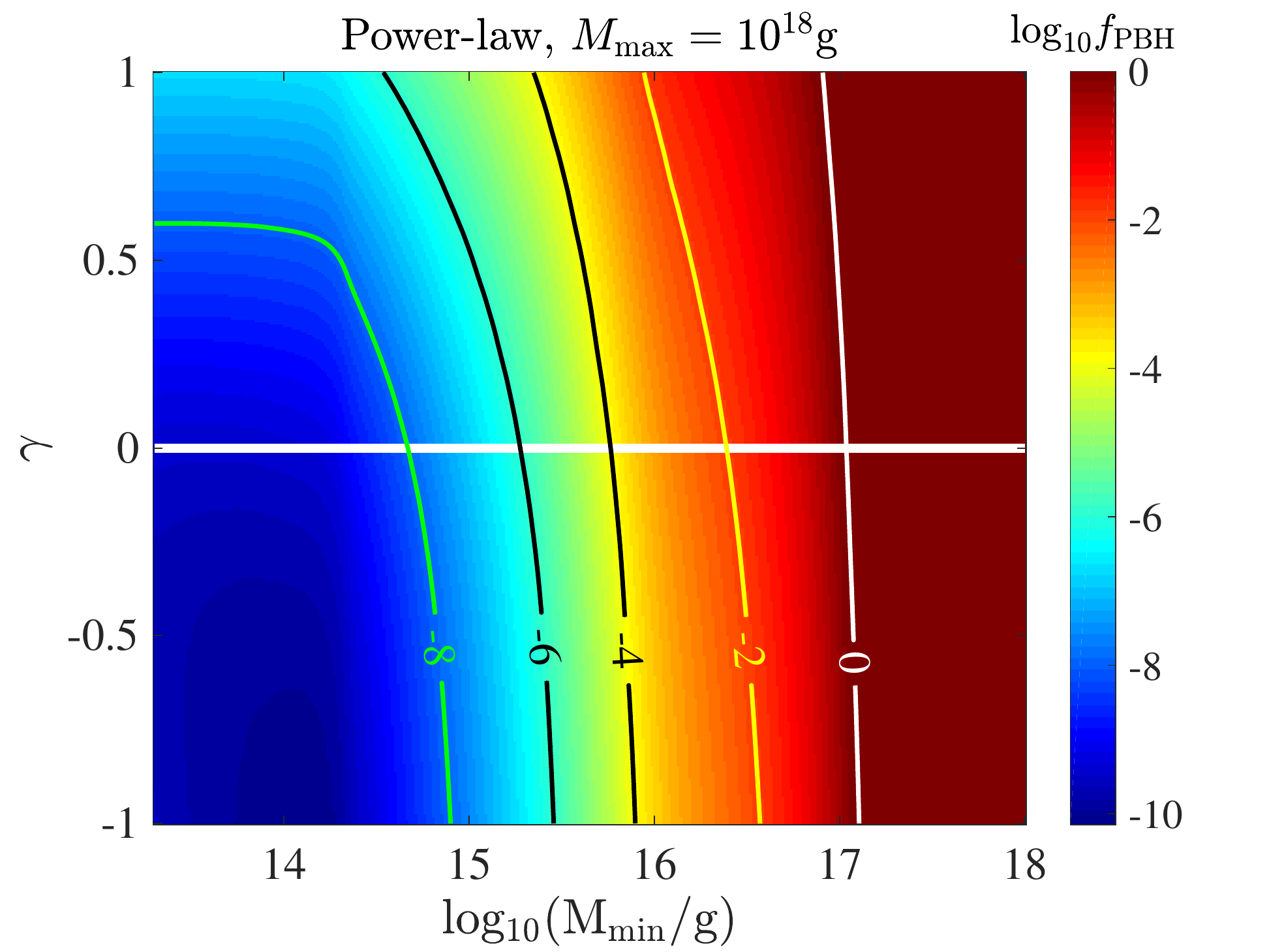}  \includegraphics[width=7.8cm]{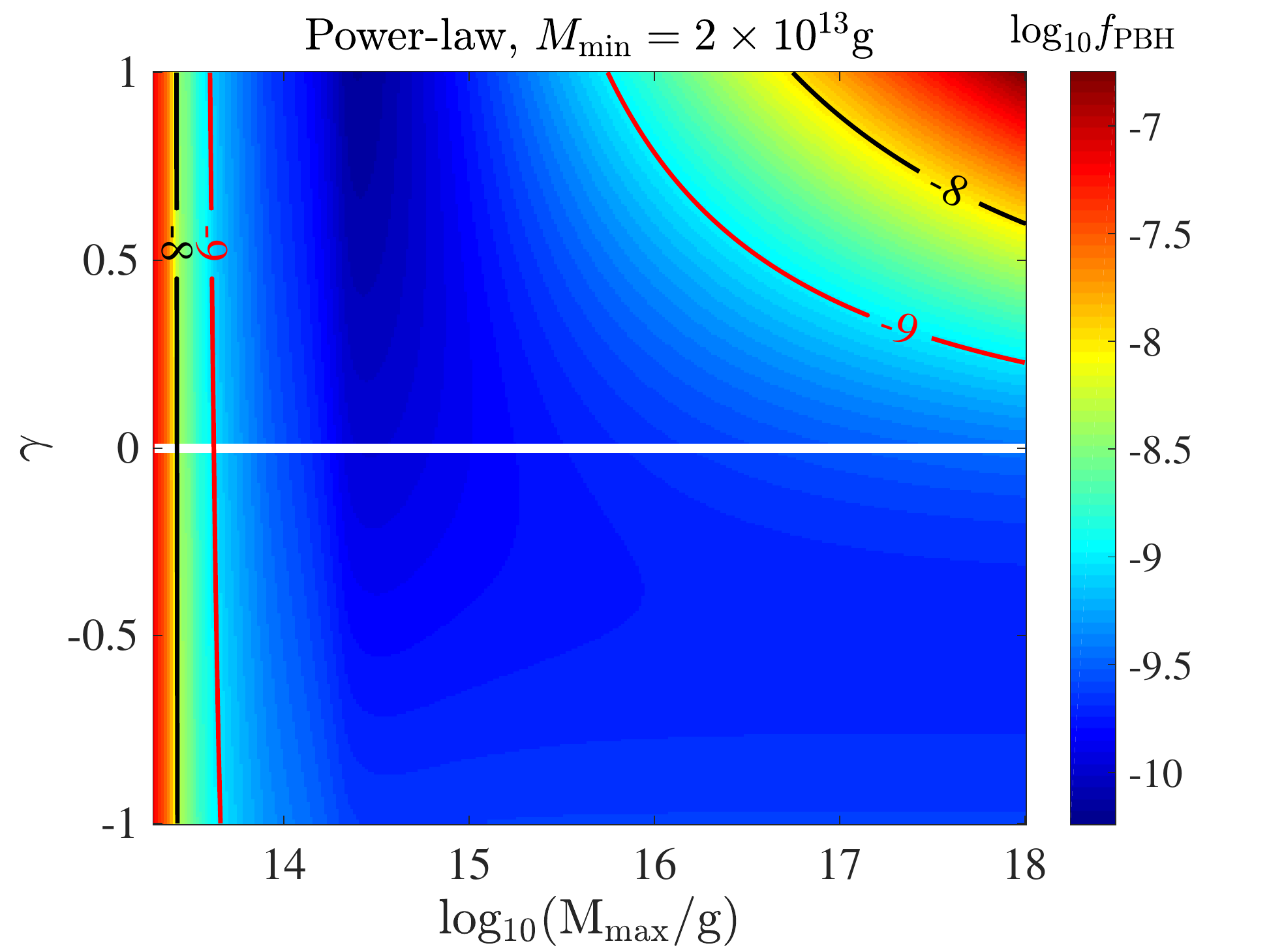}}
\subfigure[{$a_0=0.999$}]{\includegraphics[width=7.8cm]{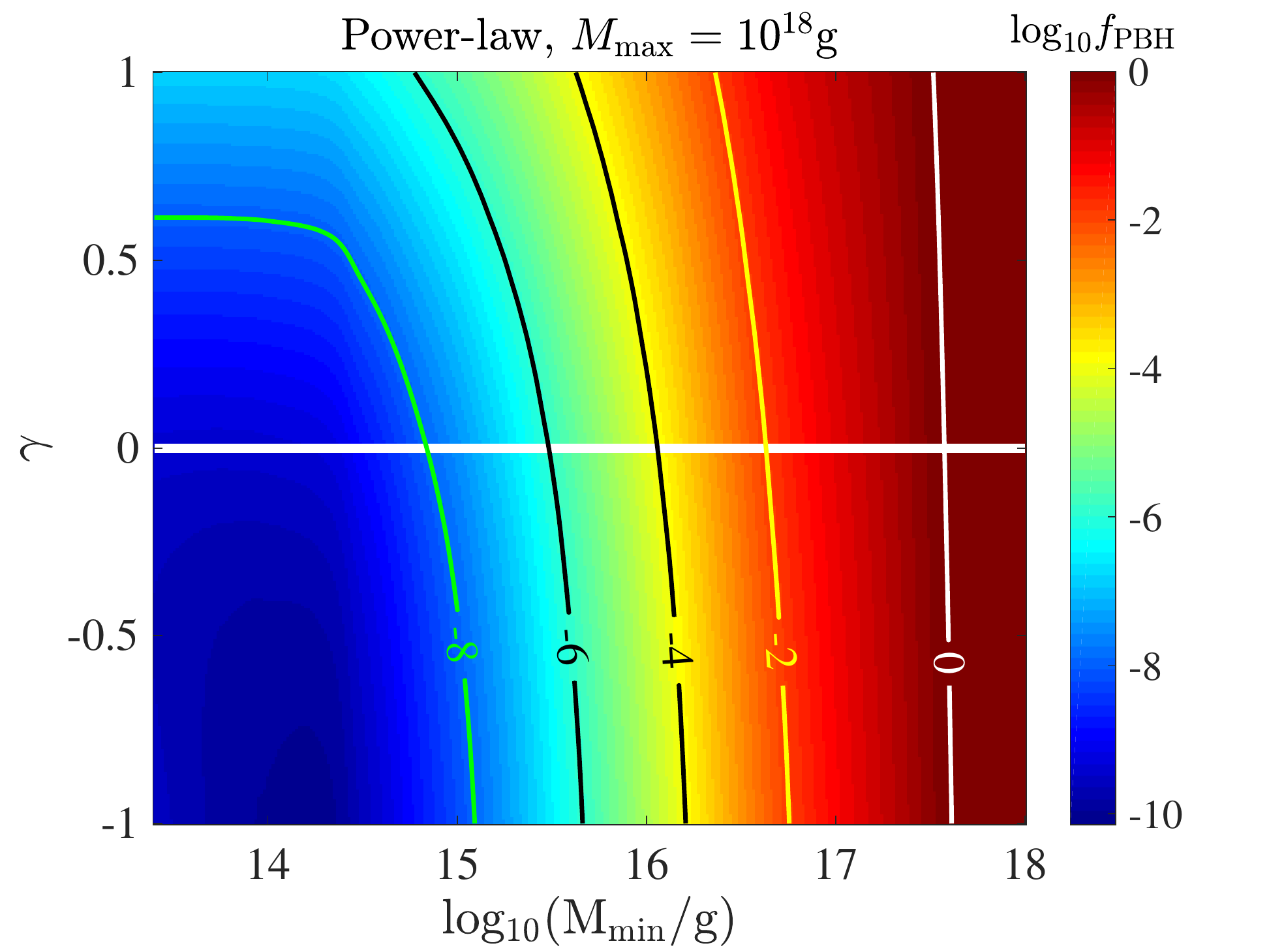}  \includegraphics[width=7.8cm]{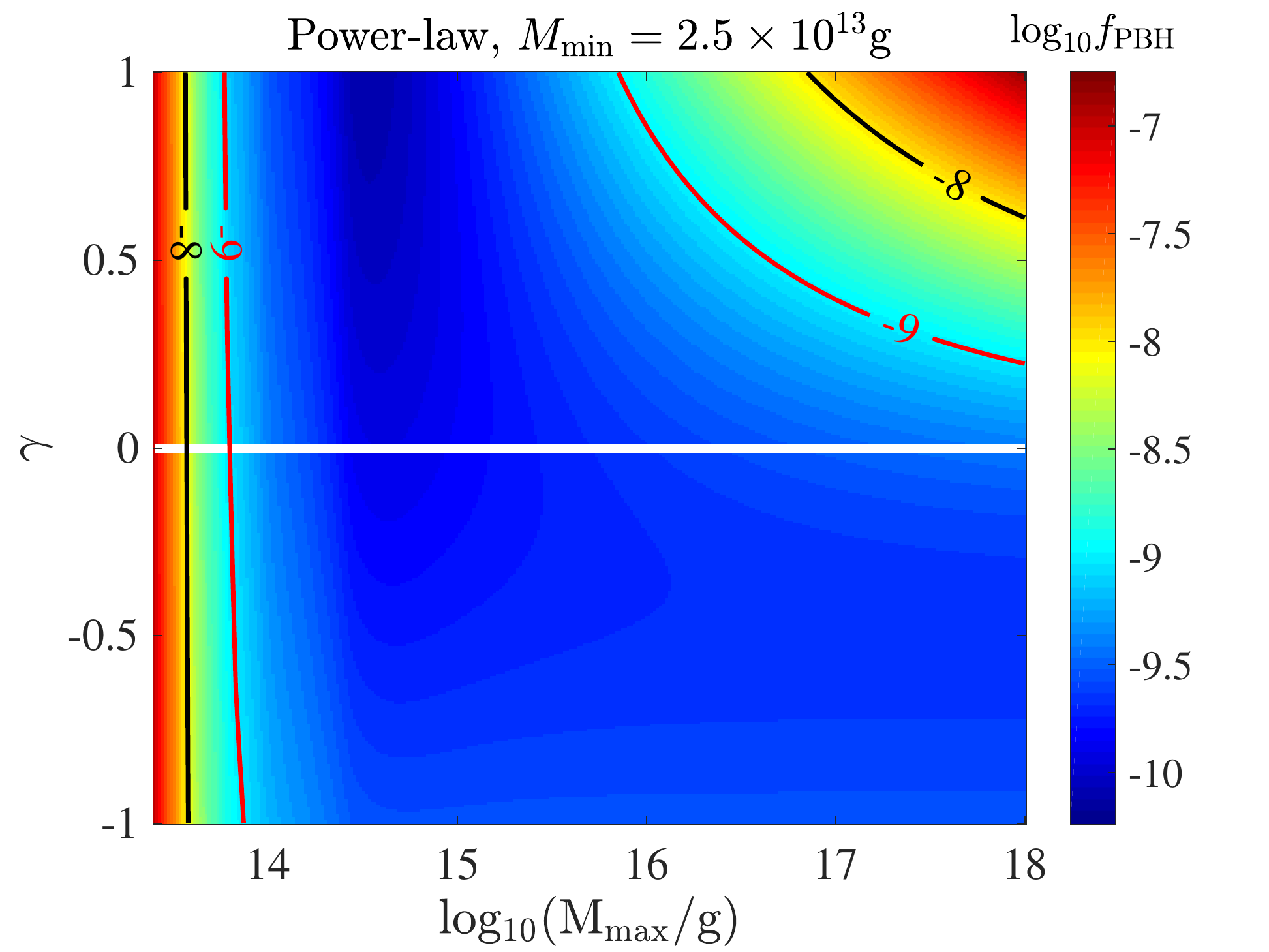}}
\caption{
21-cm upper bounds on $f_{\rm{PBH}}$ for power-law distributions with fixed $M_{\r{max}}$ (left) and $M_{\r{min}}$ (right).
The top, middle and bottom panels correspond to initial spins of $a_0=0$, $0.5$ and $0.999$ respectively.
The white horizontal lines indicate power-law index $\gamma=0$,
corresponding to PBH formation in matter dominated era and is not considered here.
White contours show regions in which PBH can account for all DM ($f_{\r{PBH}} = 1$).
}
\label{Re_2D_ds}
\end{figure*}

Figure~\ref{Re_1D} shows our 21-cm and CMB upper bounds on $f_{\r{PBH}}$ for PBHs with monochromatic and critical collapse distributions.
21-cm constraints on the log-normal and power-law models are listed in Fig.~\ref{Re_2D} and Fig.~\ref{Re_2D_ds}.
With extended mass-distributions, the number of model parameters would significantly increase the amount of computation workload.
For better computational efficiency, we use a fast re-interpretation method introduced in Refs~\cite{Carr:2017jsz,Kuhnel:2017pwq},
\be
f_{\rm{PBH}}^{-1}
\ge
\int^{10^{18}{\rm{g}}}_{2 \times 10^{13}{\rm{g}}}
 {\rm{d}} M_0
\frac
{\Psi(M_0,t_0)}
{f_{\rm{max}} (M_0)},
\label{Extended_Constraints_Re_Interpret}
\ee
where $f_{\rm{max}} (M_0)$ is the $f_{\r{PBH}}$ upper bounds on PBHs monochromatically distributed at $M_0$.
In Ref.~\cite{Carr:2017jsz} it was analytically proven that Eq. (\ref{Extended_Constraints_Re_Interpret}) is valid for $f_{\rm{max}}$ derived from fitting only one observable, such as solely comparing $T_{\r{K}}$ at redshift $z=17$ in our 21-cm constraints. 
For extended PBH distribution with multiple model parameters,
Eq. (\ref{Extended_Constraints_Re_Interpret}) is significantly faster than iteratively solving Eq. (\ref{dsdghwi6754erdfg}),
allowing us to compute $f_{\rm{PBH}}$ constraints for a high-resolution grid in the parameter space.

For all 21-cm constraints shown in this work,
we have crosschecked that results of Eq. (\ref{dsdghwi6754erdfg}) and Eq. (\ref{Extended_Constraints_Re_Interpret}) show almost identical agreement.
We also compared CMB constraints given by MCMC and Eq. (\ref{Extended_Constraints_Re_Interpret}) for two extended distribution scenarios:
the critical collapse model and a log-normal distribution with $\sigma=1$.
Unlike the 21-cm bound, 
we find that Eq. (\ref{Extended_Constraints_Re_Interpret}) does not always apply for the CMB limit.
For $M_{\r{c}} < 10^{15}\  \r{g}$,
where the evolution of PBH mass distribution cannot be ignored,
Eq. (\ref{Extended_Constraints_Re_Interpret}) tends to overestimate $f_{\rm{PBH}}$ upper bounds, 
although it agrees with {\tt CosmoMC} results for $M_{\r{c}} \ge 10^{15}\  \r{g}$.

As can be seen from Fig. (\ref{Re_1D}),
$f_{\r{PBH}}$ constraints set by Eq. (\ref{dsdghwi6754erdfg})
are more stringent than CMB by more than 2 orders of magnitude.
Spinning PBHs are typically more active than Schwarzschild ones,
therefore our constraint tightens for PBHs with higher spin.
For the conventional monochromatic model,
21-cm excludes Schwarzschild PBHs with initial mass smaller than $1.5 \times 10^{17} \ {\r{g}}$ as the dominant DM component,
whereas extreme Kerr PBHs with initial spin of $a_0=0.999$ are ruled out for initial masses below $6 \times 10^{17}\ \r{g}$.
21-cm bounds become weaker than CMB for masses below $5 \times 10^{13}\ \r{g}$, 
because it is only sensitive to ionization and gas temperature at redshift 17,
by which time these PBHs would have all vanished,
whereas CMB can probe energy injection across much higher redshifts~\cite{PhysRevD.85.043522,Cang:2020exa}.

A monochromatic distribution at $M_0$ can be described by a log-normal distribution with $M_{\r{c}}=M_0$ and $\sigma \to 0$
or a power-law distribution with $M_{\r{min}} \to M_{\r{max}}=M_0$.
Under these limits,
we find that bounds on log-normal and power-law distributions mimic the monochromatic constraints as expected.
The critical collapse distribution is very narrow and can be fitted by a sharp log-normal distribution with $\sigma=0.26$~\cite{Carr:2017jsz,Cang:2020aoo},
therefore from the right panel of \Fig{Re_1D},
one can see that bounds on this distribution is very similar to that on the monochromatic model.
As shown in the left panels of \Fig{Re_2D},
21-cm rules out log-normal parameter space roughly confined by $\left[M_\r{c}< 1.5 \times 10^{17}\ \r{g},\ \sigma >1\right]$ for $a_0=0$ 
or 
$\left[M_\r{c}< 6 \times 10^{17}\ \r{g},\ \sigma>0.65\right]$ for $a_0=0.999$.
The right panels of \Fig{Re_2D} illustrate 21-cm constraints on power-law distribution for PBHs formed during radiation dominated epoch ($\gamma=-0.5$),
for which we find that the only allowed parameter space lies roughly at $M_\r{min} > 1.5 \times 10^{17}\ \r{g}$ for $a_0=0$ and $M_\r{min} > 6 \times 10^{17}\ \r{g}$ for $a_0=0.999$.
For very wide mass distributions,
which can correspond to log-normal models with $\sigma \gg 10^2$ or power-law distributions with $M_{\r{max}} \gg 10^{18} \ \r{g}$,
the majority of PBH density will spread outside our $\left[2 \times 10^{13},\ 10^{18}\right]\ \r{g}$ mass window and cannot be efficiently probed by 21-cm or CMB,
therefore our constraints would relax for these extreme distribution scenarios.

\section{Summary}
\label{sdgsfhgvhdhh45rtgh_2}
Energy injection from evaporating primordial black holes can heat up and ionize the intergalactic medium. Accumulated energy
deposit into IGM gas leads to a potentially higher gas temperature near the early reionization epoch.
These effects damp the amplitude of 21-cm brightness temperature $T_{21}$.
Here we studied the 21-cm constraints on the abundance of both Schwarzschild and Kerr 
PBHs in $[2 \times 10^{13},\ 10^{18}]\ \r{g}$ mass window that covers four orders of magnitudes. Mass and spin evolutions
due to evaporation are accounted for, esp. at relatively low PBH mass and high spin, that leads a non-trivial radiation
energy deposit history. We considered four characteristic PBH mass distributions, in which PBHs lighter than $10^{15}$ grams typically lose a significant fraction of their masses by violent evaporation, or vanish entirely before the current age of the Universe. The corresponding
impact on IGM gas temperature and ionization evolution are numerically computed and give stringent limits on the PBH abundance and mass 
distribution by comparing with current $T_{21}$ observation data.

EDGES measures $T_{21}$ at redshift $z=17$ to be $-500^{+200}_{-500} \  {\r{mK}}$,
we place our 21-cm constraints by imposing that PBH does not raise $T_{21}$ at this redshift above $-150 \ \r{mK}$ (>99\% C.L.).
Our results show that the global 21-cm measurement provides the currently most stringent PBH constraints across our mass window.
Bounds from CMB are weaker than 21-cm by about two orders of magnitudes.
Both 21-cm and CMB limits on spinning Kerr PBHs are generally tighter than Schwarzschild ones by up to two orders of magnitudes.
For the conventional monochromatic mass distribution,
21-cm excludes Schwarzschild PBHs with initial mass below $1.5 \times 10^{17}\ \r{g}$,
extreme Kerr PBHs with reduced initial spin $a_0=0.999$ are ruled out as the dominant DM component for masses lower than $6 \times 10^{17}\ \r{g}$.

For convenience with analysis, our PBH mass and abundance limits are presented for fixed initial black hole spins.
It is likely for PBHs to have an extended initial spin distribution, and the corresponding limits can be derived 
by binning the initial spin and summing up their total heating contributions. Our PBH evolution and energy injection mainly accounts for Hawking radiation and does not include effects from black hole merger or accretion radiation. While accretion is typically small for the low PBH mass
range, mergers in principle let the PBHs grow in size, slows down their evaporation and could relax the limits from energy injection. Our PBH correction to gas temperature is computed based on a otherwise standard temperature evolution in $\Lambda$CDM cosmology, and does not include any non-standard gas cooling effects that might be hinted by the magnitude of EDGES $T_{21}$ measurement. Such effects compete with radiational heating and can be studied if confirmed by future 21-cm observations.

\medskip
{\bf Acknowledgements}
The authors thank J. Auffinger for helpful discussions of the low-energy secondary spectrum treatment in {\tt BlackHawk} v2.
J.C and Y.G. are supported by the Ministry of Science and Technology of China under the grant number 2020YFC2201601.
Y.Z.M. is supported by National Research Foundation of South Africa through grant no. 120378, 120385 and UKZN ``Big Data with Science and Society'' Research Flagship Project.

\bibliographystyle{JHEP}
\bibliography{PBH.bib}

\end{document}